\documentclass[citeautoscript,aps,prl,twocolumn,superscriptaddress,longbibliography]{revtex4-2}

\usepackage{filecontents}
\usepackage[colorlinks=true,
linkcolor=blue,
urlcolor=blue,
citecolor=blue]{hyperref}

\usepackage{physics}
\usepackage{graphicx}
\usepackage{amsmath}
\usepackage{multirow}
\usepackage{soul}
\graphicspath{{images/}}

\usepackage{soul}
\usepackage{cancel}
\usepackage{xcolor}
\usepackage{multirow}
\newcommand\Ccancel[2][black]{\renewcommand\CancelColor{\color{#1}}\cancel{#2}}
\graphicspath{{images_supplementary/}}

\newcommand{\beginsupplement}{%
	\setcounter{table}{0}
	\renewcommand{\thetable}{S\arabic{table}}%
	\setcounter{figure}{0}
	\renewcommand{\thefigure}{S\arabic{figure}}%
}

\begin{document}

\title{Natural optical activity from density-functional perturbation theory}

\author{Asier Zabalo} 
\affiliation{Institut de Ci\`encia de Materials de Barcelona (ICMAB-CSIC), Campus UAB, 08193 Bellaterra, Spain}
\author{Massimiliano Stengel}
\affiliation{Institut de Ci\`encia de Materials de Barcelona (ICMAB-CSIC), Campus UAB, 08193 Bellaterra, Spain}
\affiliation{ICREA-Instituci\'o Catalana de Recerca i Estudis Avan\c{c}ats, 08010 Barcelona, Spain}

\date{\today}

\begin{abstract}
We present an accurate and computationally efficient first-principles methodology to 
calculate the natural optical activity. Our approach is based on the long-wave density-functional 
perturbation theory and includes self-consistent field (SCF) terms naturally in the formalism, which are found to be of crucial importance. 
The final result is expressed exclusively in terms of response functions to
uniform field perturbations and avoids troublesome summations over empty states.
Our strategy is validated by computing the 
natural optical activity tensor in representative chiral crystals (trigonal Se, $\alpha$-HgS and $\alpha$-SiO$_2$) and molecules (C$_4$H$_4$O$_2$), 
finding excellent agreement with experiment and previous theoretical calculations.
\end{abstract}

\maketitle
Natural optical activity (NOA) refers to the first-order spatial dispersion of the macroscopic dielectric tensor~\cite{landau1984electrodynamics}.
Empirically, it manifests as optical rotation (OR), which is a property of certain structures to rotate the plane of the 
polarization of light that travels through them \cite{RevModPhys.9.432,mason1982molecular}; at difference with
the Faraday effect, NOA is reciprocal and doesn't require magnetism to be present~\cite{melrose1991electromagnetic}.
It was first measured in quartz crystals back in 1811 by Arago, and 
historically, most of the studied optically active materials turned 
out to be chiral. 
In fact, chirality is a sufficient but not necessary
condition for NOA to be present, as 
optically active achiral systems also exist \cite{doi:10.1021/cm980140w}.
Since its discovery, natural optical activity
has been attracting increasing research interest, and reliable experimental measurements
now exist for many materials, both in molecular 
\cite{stephens2001calculation,doi:10.1063/1.1436466,doi:10.1063/1.1477925,doi:10.1080/002689797171436,mort2005magnitude} 
and crystalline form \cite{doi:10.1063/1.433207,Ades_75,stadnick,https://doi.org/10.1002/pssb.2220680247,PhysRevLett.5.500,doi:10.1063/1.1653691,https://doi.org/10.1107/S0108767386098732}.

Parallel to the experiments, there have been considerable advances in the 
theoretical understanding of optical rotation as well
\cite{RevModPhys.9.432,doi:10.1063/1.433207,barron2009molecular,PhysRevB.48.1384,PhysRevLett.69.379}.
\textit{Ab-initio} methods like Hartee-Fock (HF), \cite{doi:10.1080/002689797171436} coupled-cluster (CC) \cite{doi:10.1063/1.1772352} and density functional theory (DFT) \cite{stephens2001calculation,doi:10.1063/1.1647515,doi:10.1063/1.1436466} have 
recently become popular in the context of NOA.
While most of the available literature is about small molecules, notable attempts at calculating optical activity in solids do exist.
It is worth mentioning, for example, 
the pioneering works by Zhong, Levine, Allan and Wilkins, \cite{PhysRevB.48.1384,PhysRevLett.69.379} based  
on a numerical long-wavelength expansion of the electromagnetic response function.
Later, Malashevich and Souza \cite{PhysRevB.82.245118} and Pozo and Souza \cite{pozo2022multipole}
derived analytical expressions for the NOA, thus reviving the interest in the field;
their formalism has been implemented very recently within an \textit{ab initio} context 
\cite{PhysRevB.107.045201}.
The agreement between theory and experiment achieved in these works
is quite good, e.g., for trigonal Se \cite{pozo2022multipole,PhysRevB.82.245118}, $\alpha$-quartz \cite{PhysRevB.48.1384,PhysRevLett.69.379} and trigonal Te \cite{PhysRevB.97.035158}.

In spite of the remarkable progress, however, a systematic, first-principles-based and 
computationally efficient methodology to compute the NOA has not 
been established yet.
The first issue concerns the treatment of the self-consistent (SCF) fields. These were 
accounted for in Ref.~\cite{PhysRevLett.76.1372} and found to be of crucial importance,
but the numerical differentiations with respect to the wave vector ${\bf q}$ that were
used therein have limited a widespread application of their method.
The existing analytical expressions~\cite{PhysRevB.82.245118,pozo2022multipole} for the NOA are, 
in principle, better suited to an \emph{ab initio} implementation \cite{PhysRevB.107.045201}, but the SCF contributions are 
systematically neglected therein.
Another disadvantage with the existing techniques lies in that they require 
cumbersome sums over
empty states; this introduces an additional
potential source of error, as the convergence with respect to the number of bands tends
to be slow.
There are additional technical subtleties that have not been considered in the context
of the NOA, for example regarding the correct treatment of the current-density response
in presence of nonlocal pseudopotentials~\cite{PhysRevB.98.075153}.
It is unquestionable that the current limitations rule out the study of many systems of outstanding interest
(e.g., electrotoroidic compounds \cite{PhysRevB.87.195111,naumov2004unusual}), which are hardly accessible 
to the currently available schemes.

Here we present, within the framework of first-principles long-wave density 
functional perturbation theory (DFPT), a method 
to calculate the natural optical activity that overcomes the aforementioned
limitations and is equally valid for molecules and extended solids.
Building on Ref. \cite{PhysRevX.9.021050}, we express the natural optical activity tensor as the first-order 
spatial dispersion (i.e., derivative with respect to the wave vector $\mathbf{q}$) of the macroscopic 
dielectric function. 
Crucially, the capabilities of the recently implemented long-wave module \cite{doi:10.1063/1.5144261} of {\sc abinit} \cite{GONZE2020107042,GONZE20092582} allow for an efficient calculation by combining response functions that are 
already available in the code (e.g., $\mathbf{k}$-derivatives, electric and orbital magnetic field 
perturbations).
This way, summations over excited states are entirely avoided, and the effect of local fields 
is automatically included
without the need for an ad-hoc treatment.
We validate our methodology by computing the NOA tensor
for well known chiral structures, including trigonal crystals (Se, $\alpha$-HgS and $\alpha$-SiO$_2$) and
the C$_4$H$_4$O$_2$ molecule. 
Our numerical results show fast convergence with respect to the main computational parameters, 
and are in excellent agreement with experiment and earlier theoretical calculations.

Our starting point is the double Fourier transform in frequency $\omega$ and wave vector $\mathbf{q}$ of the permittivity function, 
$\epsilon_{\alpha\beta}(\omega,\mathbf{q})$.  By expanding $\epsilon_{\alpha\beta}(\omega,\mathbf{q})$ in powers of the wave vector 
$\mathbf{q}$, around $\mathbf{q=0}$, we obtain
\begin{equation}
\epsilon_{\alpha\beta}(\omega,\mathbf{q})=
\epsilon_{\alpha\beta}(\omega,\mathbf{q=0})
+iq_\gamma\eta_{\alpha\beta\gamma}(\omega)
 +\dots,
\end{equation}
where $\eta_{\alpha\beta\gamma}(\omega)$ is the natural optical activity tensor
\cite{landau1984electrodynamics}.
(From now on, we adopt Einstein summation conventions for the Cartesian indices $\alpha\beta\gamma$.) 
In absence of dissipation (i.e., in the transparent regime), 
$\epsilon_{\alpha\beta}(\omega,\mathbf{q})$ is a $3\times 3$ Hermitian matrix, which 
at ${\bf q=0}$ becomes real symmetric in crystals with time-reversal (TR) symmetry.
The frequency-dependent natural optical activity tensor is then
also real and satisfies $\eta_{\alpha\beta\gamma}(\omega)=
-\eta_{\beta\alpha\gamma}(\omega)$, 
which means that only 9 of the 27 components of $\eta_{\alpha\beta\gamma}$ are independent.
As a consequence, $\eta_{\alpha\beta\gamma}$ is often rearranged into the second-rank \textit{gyration} or \textit{gyrotropic} tensor, $g_{\alpha\beta}$,
\cite{landau1984electrodynamics}
\begin{equation}\label{Eq_g_gyration}
g_{\alpha\beta}(\omega)=\frac{1}{2}\epsilon_{\gamma\delta\alpha}
\eta_{\gamma\delta\beta}(\omega),
\end{equation}
where $\epsilon_{\gamma\alpha\delta}$ is the Levi-Civita symbol. 
Assuming a crystal structure with the point group 32 (trigonal Se, $\alpha$-HgS and $\alpha$-SiO$_2$ belong to this 
crystal class), and considering that the optical axis is oriented along the $z$ Cartesian direction \cite{PhysRevB.48.1384},
\begin{equation}
\boldsymbol{g}(\omega)=\begin{pmatrix}
g_{11}(\omega) &0& 0\\
0& g_{11}(\omega)&0\\
0&0&g_{33}(\omega)
\end{pmatrix},
\end{equation}
where $g_{11}=\eta_{231}$ and $g_{33}=\eta_{123}$. The optical rotatory
power $\rho$ is then given by \cite{PhysRevB.48.1384}
\begin{equation}\label{Eq_Optical_Rot_Power}
\rho(\omega)=\frac{\omega^2}{2c^2}g_{33}(\omega),
\end{equation}
where $c$ is the speed of light. 
In this work, we shall focus on the $\omega\rightarrow 0$ limit, where
the components of both $\boldsymbol{g}$ and $\boldsymbol{\eta}$ tend to a finite constant,
\begin{equation}\label{Eq_eta}
\eta_{\alpha\beta\gamma}=\eta_{\alpha\beta\gamma}(\omega \rightarrow 0),
 \qquad g_{\alpha \beta} = g_{\alpha \beta}(\omega \rightarrow 0).
\end{equation}
At leading order in the frequency, this yields a rotatory power of
\begin{equation}\label{Eq_bar_rho}
\rho(\omega) \simeq  (\hbar \omega)^2 \bar{\rho},
\qquad \bar{\rho} =\frac{ g_{33} }{2(\hbar c)^2},
\end{equation}
where $\hbar$ is the reduced Planck constant. The constant $\bar{\rho}$ is usually expressed in the units of deg/[mm (eV)$^2$] and can be directly compared to
experimental measurements.

To make further progress,
we shall express the dielectric function in the low-frequency limit as a second derivative of the ground state 
energy with respect to two spatially modulated electric fields ($\boldsymbol{\mathcal{E}}$) \cite{PhysRevB.55.10355}
\begin{equation}
\epsilon_{\alpha\beta}(\mathbf{q})=\delta_{\alpha\beta}-
\frac{4\pi}{\Omega}E_\mathbf{q}^{\mathcal{E}_\alpha \mathcal{E}_\beta}.
\end{equation}
This allows us to write the natural optical activity tensor as the first derivative of $\epsilon_{\alpha\beta}(\mathbf{q})$ with respect to 
$q_\gamma$,
\begin{equation}\label{Eq_eta_omega}
\eta_{\alpha\beta\gamma}=-\frac{4\pi}{\Omega}\text{Im}
E_\gamma^{\mathcal{E}_\alpha \mathcal{E}_\beta}, \qquad
 E_\gamma^{\mathcal{E}_\alpha \mathcal{E}_\beta} = \frac{\partial E_\mathbf{q}^{\mathcal{E}_\alpha \mathcal{E}_\beta}}{\partial q_\gamma}\Big|_{\bf q=0},
\end{equation}
where $\Omega$ is the volume of the unit cell.
By virtue of the ``$2n+1$'' theorem~\cite{PhysRevX.9.021050}, $E_\gamma^{\mathcal{E}_\alpha \mathcal{E}_\beta}$ can be written 
in terms of uniform-field response functions, which are already 
available in public first-principles packages like {\sc abinit}.
More specifically, we find
\begin{equation}\label{Eq_E_gamma_EE}
E_\gamma^{\mathcal{E}_\alpha \mathcal{E}_\beta}=E_{\text{elst},\gamma}^{\mathcal{E}_\alpha \mathcal{E}_\beta} + 2s\int_\text{BZ} [d^3k]
E_{\mathbf{k},\gamma}^{\mathcal{E}_\alpha \mathcal{E}_\beta},
\end{equation}
where $s=2$ is the spin multiplicity, and the shorthand notation $[d^3k] = {\Omega}/{(2\pi)^3} d^3k$ is used
for the Brillouin-zone (BZ) integral. (We assume
that the system under study is a time-reversal (TR) symmetric insulator.)
The electrostatic (elst) term is defined as 
\begin{equation}\label{Eq_E_elst}
E_{\text{elst},\gamma}^{\mathcal{E}_\alpha \mathcal{E}_\beta}=\int_\Omega \int 
n^{\mathcal{E}_\alpha}(\mathbf{r})K_\gamma(\mathbf{r,r'})n^{\mathcal{E}_\beta}d^3rd^3r',
\end{equation}
where 
$n^{\mathcal{E}_\beta}$ is the first-order charge density response to $\mathcal{E}_\beta$, 
and $K_\gamma(\mathbf{r,r'})$ is the first $\mathbf{q}$-derivative of the Hartree exchange and correlation (Hxc) kernel. 
The wave function term of Eq.~(\ref{Eq_E_gamma_EE}), in turn, can be written as
\begin{equation}\label{E_calligraphic_symbols}
\begin{split}
E_{\mathbf{k},\gamma}^{\mathcal{E}_\alpha \mathcal{E}_\beta}=&
\mathcal{X}_{\mathbf{k}}^{\mathcal{E}_\alpha k_\gamma\mathcal{E}_\beta}
+\mathcal{Y}_{\mathbf{k}}^{\mathcal{E}_\alpha \mathcal{E}_\beta k_\gamma }
+\mathcal{Y}_{\mathbf{k}}^{ k_\gamma \mathcal{E}_\alpha \mathcal{E}_\beta}
\\
&+\mathcal{W}^{\alpha,\beta \gamma}_{\bf k} + \left(\mathcal{W}^{\beta,\alpha \gamma}_{\bf k} \right)^*,
\end{split}
\end{equation} 
We shall explain Eq.~(\ref{E_calligraphic_symbols}) term by term in the following.

For three generic perturbations, $\lambda_1$, $\lambda_2$ and $\lambda_3$, the calligraphic 
symbols in the first line are defined as 
\begin{equation}
\mathcal{X}_{\mathbf{k}}^{\lambda_1\lambda_2\lambda_3}=\sum_m\bra*{u_{m\mathbf{k}}^{\lambda_1}}
\hat{\mathcal{H}}^{\lambda_2}_\mathbf{k}\ket*{u_{m\mathbf{k}}^{\lambda_3}}
\end{equation}
and
\begin{equation}
\mathcal{Y}_{\mathbf{k}}^{\lambda_1\lambda_2\lambda_3}=
-\sum_{mn}\bra*{u_{m\mathbf{k}}^{\lambda_1}}\ket*{u_{n\mathbf{k}}^{\lambda_3}}
\bra*{u_{n\mathbf{k}}^{(0)}}\hat{\mathcal{H}}_\mathbf{k}^{\lambda_2}\ket*{u_{m\mathbf{k}}^{(0)}}.
\end{equation}
(The band indices $m,n$ run over the occupied states only.)
Here, $\ket*{u_{m\mathbf{k}}^\lambda}$ are the first-order wave functions and
the first-order calligraphic Hamiltonian is given by $\hat{\mathcal{H}_\mathbf{k}^\lambda}=\hat{H}^\lambda_\mathbf{k}+\hat{V}^\lambda$, 
where $\hat{H}^\lambda_\mathbf{k}$ is the external perturbation and $\hat{V}^\lambda$ is the 
self-consistent field (SCF) potential response. Note that $\hat{\mathcal{H}}_\mathbf{k}^{k_\gamma} = \hat{{H}}_\mathbf{k}^{k_\gamma}$
as there is no SCF contribution to the derivative in $\mathbf{k}$-space, and $\hat{\mathcal{H}}_\mathbf{k}^{\mathcal{E}_\alpha} = \hat{V}^{\mathcal{E}_\alpha}$
in the above equations since the ``external potential'' is a purely cross-gap operator in the electric-field case~\cite{PhysRevX.9.021050}.

The third line is defined as
\begin{equation}
\mathcal{W}^{\alpha,\beta \gamma}_{\bf k} = \sum_m i
\bra*{u_{m\mathbf{k}}^{\mathcal{E}_\alpha}}\ket*{u_{m\mathbf{k},\gamma}^{A_\beta}},
\end{equation}
where $\ket*{u_{m\mathbf{k},\gamma}^{A_\beta}}$ indicates the wave function response to an
electromagnetic vector potential at 
first order in the modulation vector ${\bf q}$. (See Sec. IV of Ref. \cite{supplemental_prl} for more details.) 
We can write $\mathcal{W}$
as a sum of two contributions that are, respectively, symmetric ($\mathcal{S}_\mathbf{k}^{\alpha,\beta\gamma}$) and
an antisymmetric ($\mathcal{A}_\mathbf{k}^{\alpha,\beta\gamma}$) with respect to 
$\beta\leftrightarrow\gamma$ exchange, 
\begin{equation}
\mathcal{W}^{\alpha,\beta \gamma}_{\bf k} = \mathcal{S}_\mathbf{k}^{\alpha,\beta\gamma} + \mathcal{A}_\mathbf{k}^{\alpha,\beta\gamma}.
\end{equation}
These objects are given by
\begin{equation}\label{Eq_S}
\mathcal{S}_\mathbf{k}^{\alpha,\beta\gamma}=\frac{i}{2} \sum_m\bra*{u_{m\mathbf{k}}^{\mathcal{E}_\alpha}}\ket*{\partial_{\beta\gamma}^2u_{m\mathbf{k}}^{(0)}}
\end{equation}
and
\begin{equation}\label{Eq_A}
\mathcal{A}_\mathbf{k}^{\alpha,\beta\gamma}=\frac{1}{2}\sum_m \epsilon_{\delta \beta\gamma} \bra*{u_{m\mathbf{k}}^{\mathcal{E}_\alpha}}\ket*{u_{m\mathbf{k}}^{B_\delta}}.
\end{equation}
In Eq. (\ref{Eq_S}), $\partial^2_{\beta\gamma}$ represents a second derivative in 
$\mathbf{k}$-space. The 
$\ket*{\partial_{\beta\gamma}^2 u_{m\mathbf{k}}^{(0)}}$
functions in 
$\mathcal{S}$ are the well known $d^2/dk_\beta dk_\gamma$ wave functions 
\cite{PhysRevX.9.021050,PhysRevB.105.094305,PhysRevB.84.094304};
 whereas in 
Eq.~(\ref{Eq_A}),
$\ket*{u_{m\mathbf{k}}^{B_\delta}}$ is the 
wave function response to a uniform orbital magnetic field, $B_\delta$, 
as defined in Refs. 
\cite{PhysRevB.81.205104,PhysRevB.84.064445}.

For finite systems, the above theory nicely recovers the established formulas that are used in quantum 
chemistry calculations (more details can be found in Sec. VI of Ref. \cite{supplemental_prl}). 
Our formulation, however, presents many crucial advantages. First, 
 Eq.~(\ref{Eq_E_gamma_EE}) has been derived within a DFPT framework,
and hence avoids the cumbersome summations over unoccupied states that are 
required by other methods.
Second, all contributions to Eq.~(\ref{E_calligraphic_symbols}) are
individually independent of the choice of the origin, and equally
valid for both molecules and extended crystals; this implies that our formulas
are free of cancellation errors due to incomplete basis sets.
Third, all the aforementioned terms are
independent of the choice of the wave function gauge by construction, 
as they are all expressed as parametric derivatives (with respect to ${\bf q}$)
of multiband gauge-invariant quantities.
Fourth, the treatment of the current-density response in presence of
nonlocal pseudopotentials complies with the prescriptions of Ref.~\cite{PhysRevB.98.075153}.
Finally, and most importantly, SCF terms naturally appear in our formalism, both 
directly in $E_\text{elst}$ and $\mathcal{Y}$ (both terms vanish if local fields 
are neglected), and indirectly in the other terms 
via the first-order wave functions 
$\ket*{u_{m\mathbf{k}}^{\mathcal{E}_\alpha}}$ 
(see Sec. V. of Ref. \cite{supplemental_prl}).
A natural question to ask at this point is whether Eq. (\ref{E_calligraphic_symbols}) is unique, or
whether there other combinations of the same ingredients that yield the same result.
Two inequivalent definitions of $E_{\mathbf{k},\gamma}^{\mathcal{E}_\alpha^*\mathcal{E}_\beta}$
can, at most, differ by a vanishing Brillouin-zone integral; so the question boils down to
asking whether we can combine the individual pieces in Eq. (\ref{E_calligraphic_symbols}) in such a way that the result
is the total $\mathbf{k}$-derivative of some function $f({\bf k})$.
An obvious choice for $f({\bf k})$ consists in identifying it with the $\mathbf{k}$-derivative of the 
macroscopic dielectric tensor.
Indeed, by applying the ``$2n+1$'' theorem to the stationary 
expression~\cite{PhysRevA.52.1096,PhysRevA.52.1086,PhysRevB.55.10355} for
$E_{\mathbf{k},{\bf q=0}}^{\mathcal{E}_\alpha^*\mathcal{E}_\beta}$, we find
\begin{equation}\label{Eq_E_wf_functional}
\begin{split}
\frac{\partial E_{\mathbf{k,q}}^{\mathcal{E}_\alpha^*\mathcal{E}_\beta}}{\partial k_\gamma}\Big|_{\mathbf{q=0}}=& 
\mathcal{X}_\mathbf{k}^{\mathcal{E}_\alpha\mathcal{E}_\beta k_\gamma}
+\mathcal{X}_\mathbf{k}^{k_\gamma\mathcal{E}_\alpha\mathcal{E}_\beta}
+\mathcal{X}_\mathbf{k}^{\mathcal{E}_\alpha k_\gamma \mathcal{E}_\beta}\\
&+\mathcal{Y}_\mathbf{k}^{\mathcal{E}_\alpha\mathcal{E}_\beta k_\gamma}
+\mathcal{Y}_\mathbf{k}^{\mathcal{E}_\alpha k_\gamma\mathcal{E}_\beta}
+
\mathcal{Y}_\mathbf{k}^{k_\gamma\mathcal{E}_\alpha\mathcal{E}_\beta}\\
&+2\mathcal{S}_\mathbf{k}^{\alpha,\beta\gamma}
+2\Big(\mathcal{S}_\mathbf{k}^{\beta,\alpha\gamma}\Big)^*.
\end{split}
\end{equation}
Then, by subtracting the latter expression from Eq. (\ref{E_calligraphic_symbols}), we obtain
another equally valid formula for the NOA,
\begin{equation}\label{Eq_E_wf_functional_tilde}
\begin{split}
\Big[E_{\mathbf{k},\gamma}^{\mathcal{E}_\alpha\mathcal{E}_\beta}\Big]'=&-\Big(
\mathcal{X}_\mathbf{k}^{\mathcal{E}_\alpha\mathcal{E}_\beta  k_\gamma}
+\mathcal{X}_\mathbf{k}^{k_\gamma\mathcal{E}_\alpha\mathcal{E}_\beta}
+\mathcal{Y}_\mathbf{k}^{\mathcal{E}_\alpha k_\gamma\mathcal{E}_\beta}\Big)\\
&-\mathcal{W}^{\alpha,\gamma\beta}_{\bf k} - \left(\mathcal{W}^{\beta,\gamma\alpha}_{\bf k}
\right)^*.
\end{split}
\end{equation}
Numerical tests confirm the consistency of Eq. (\ref{E_calligraphic_symbols}) and (\ref{Eq_E_wf_functional_tilde}) to a very high 
degree of accuracy.
We therefore conclude that
Eq.~(\ref{Eq_E_gamma_EE}) is not unique; 
on the contrary, there are infinite possible 
definition of the gyrotropy tensor, differing from our Eq. (\ref{Eq_E_gamma_EE}) by a 
dimensionless constant times Eq.~(\ref{Eq_E_wf_functional}).

This arbitrariness can be regarded a direct consequence of the \emph{electromagnetic} (EM) gauge freedom. 
Indeed, the last lines in both Eq. (\ref{E_calligraphic_symbols}) and Eq. (\ref{Eq_E_wf_functional_tilde}) have the physical 
meaning of Berry curvatures in the parameter space spanned by a uniform magnetic field (${\bf B}$) and an electric field.
Such curvatures are, as we said, insensitive to the choice of the coordinate origin and the wave function gauge. This
result was achieved by expressing the ${\bf B}$-field response function in a cell-periodic form, consistent with the
density-operator theory of Essin et al. \cite{PhysRevB.81.205104}.  
Notwithstanding these undeniable advantages, the aforementioned Berry curvatures retain an inherent 
dependence on the EM-gauge~\cite{doi:10.1063/5.0135923}.
More specifically, the symbol $\mathcal{W}^{\alpha,\beta\gamma}$ is expressed in a Landau gauge 
where the $\beta$ component of the ${\bf A}$-field increases linearly along $\gamma$; so when going
from Eq. (\ref{E_calligraphic_symbols}) to Eq. (\ref{Eq_E_wf_functional_tilde}) we have essentially switched between 
two Landau gauges in the last term, and collected the leftovers in the form of $\mathcal{X}$ and $\mathcal{Y}$.
(It is, of course, possible to define a third variant of Eq. (\ref{E_calligraphic_symbols}), where the contribution of 
$\mathcal{S}$ cancels out, at the expense of having a slightly longer list of $\mathcal{X}$- and $\mathcal{Y}$-symbols.)
%
Ideally, one would like to exploit this freedom to obtain a physically intuitive separation
between well-defined (and possibly individually measurable) physical effects; whether such a choice 
exists is an interesting open question, which we shall defer to a later work.

\begin{table}
	\begin{ruledtabular}
		\caption{Calculated independent components of the gyration tensor (in bohr) and the optical rotatory power $\bar{\rho}$ defined in Eq. (\ref{Eq_bar_rho}) (in deg/[mm (eV)$^2$] units). Values in brackets are computed neglecting the SCF terms.}
		\begin{tabular}{cccc}
			&$g_{11}$ & $g_{33}$ & $\bar{\rho}$\\ \hline
			Se&-1.307 (-1.547)&-1.913 (-0.458)  & -74.5 (-17.8)\\
			$\alpha$-HgS&0.775 (0.554)&-1.861 (-1.274) & -72.5 (-49.6)\\
			$\alpha$-SiO$_2$&-0.071 (-0.001)& 0.125 (0.019)& 4.9 (0.7)
		\end{tabular}
	\label{Table_rot_power}
	\end{ruledtabular}
\end{table}

Our first principles calculations are performed with the
open-source {\sc abinit} \cite{GONZE20092582,GONZE2020107042} package. (Details of the computational parameters 
are provided in Sec. I of Ref. \cite{supplemental_prl}.)
Overall, our approach displays a remarkably fast convergence with respect to the main computational 
parameters (plane-wave energy cutoff and number of $\mathbf{k}$ points, see Ref. \cite{supplemental_prl}, Sec. III). 
In Table \ref{Table_rot_power} we show the converged numerical values for the independent components of the 
gyration tensor and the optical rotatory power in our test set of trigonal crystals: 
trigonal Se, $\alpha$-HgS 
and $\alpha$-SiO$_2$ (numerical values in brackets are obtained neglecting SCF terms). 
Our results 
are in fairly good agreement with literature
values, even if a scissor operator was applied in 
Ref. \cite{PhysRevLett.76.1372} to correct the LDA band gap.
(Trigonal Se is an interesting exception, in that 
Ref. \cite{PhysRevLett.76.1372} reports an opposite sign to ours for the non-SCF value of the $g_{33}$ component; 
although the reason for this discrepancy is unclear, we remain confident in the accuracy of our results, 
as other values nicely agree with ours in both magnitude and sign.)
Overall, our results 
confirm the crucial importance of local-field SCF contributions, consistent 
with the conclusions of Ref. \cite{PhysRevLett.76.1372}.

\begin{table}[b!]
	\begin{ruledtabular}
		\caption{Comparison between LDA and GGA for the independent components of the gyration tensor for Se, 
			$\alpha$-HgS and $\alpha$-SiO$_2$, for different structures. In the Structure column, ``exp'' refers to the experimental structure, while Se (LDA) means that the structure was relaxed with LDA, for example.}
		\begin{tabular}{crrrr}
			Structure& \multicolumn{2}{c}{$g_{11}$ (bohr)}&\multicolumn{2}{c}{$g_{33}$ (bohr)}  \\ 
			&LDA&GGA&LDA&GGA\\ \hline
		Se(exp) &-1.306&-1.301&-1.910&-1.329\\
		Se(GGA) &-1.408&-1.431&-1.802&-1.216\\ \hline
		$\alpha$-HgS (LDA) & 0.775&0.663&-1.861&-1.645\\
		$\alpha$-HgS (GGA) & -0.716&-0.692 &-0.065&-0.065\\\hline
		$\alpha$-SiO$_2$ (LDA) &-0.071 &-0.071& 0.125& 0.125\\
		$\alpha$-SiO$_2$ (GGA) & -0.085&-0.085 &0.168& 0.167
		\end{tabular}
	\label{Table_GGA_LDA_2}
	\end{ruledtabular}
\end{table}

Given the large impact of SCF fields on the results, we decided to repeat our calculations 
within the PBE \cite{PhysRevLett.77.3865} parametrization of the GGA.
The corresponding values are reported in Table \ref{Table_GGA_LDA_2}.
(Further details can be found in Ref. \cite{supplemental_prl}, Sec. II.)
Interestingly, for a given crystal structure the choice between LDA and GGA seems to have 
a relatively small influence on the calculated coefficients, except for the $g_{33}$ component
of Se where such deviation reaches $\sim$50\%.  
Conversely, the structural parameters do appear to have a significant impact on the final result.
To account for this fact, we have tested various models for the crystal structure, either 
using the experimental (exp) one, or relaxed to mechanical equilibrium (either within LDA
or GGA).
Our analysis shows that the fundamental gap depends on the volume
of the unit cell, and such a dependence has a strong impact on the
calculated $\boldsymbol{g}$-tensor components. 
For example, in the LDA equilibrium structure of Se the electronic band gap is so 
small that we were unable to converge $g_{11}$ and $g_{33}$ to 
meaningful values.
While GGA displays the usual overcorrection of the equilibrium volume, 
it yields results that are in much closer agreement with the experiment.
It is also interesting to note that the NOA, unlike other linear-response properties
(e.g. the dielectric tensor), has a nontrivial dependence on the structure (and
hence on the amplitude of the gap). 
The final result originates from the mutual cancellation of several terms,
not all of which are expected to diverge in the metallic limit. This means that 
some components of ${\boldsymbol{g}}$ may change rather dramatically with structure, while 
others remain essentially unaltered (see Sec. II of Ref. \cite{supplemental_prl} for more details). 	

We now focus on the isolated molecule C$_4$H$_4$O$_2$.
\begin{table}
	\begin{ruledtabular}
		\caption{Calculated independent components of the gyration tensor times the volume of the simulation cell ($\Omega$) for C$_4$H$_4$O$_2$. Values are given in Hartree atomic units.}
		\begin{tabular}{lccccc}
			&$\Omega g_{11}$ & $\Omega g_{22}$ & $\Omega g_{33}$&$\frac{\Omega}{2}(g_{12}+g_{21})$& $\beta$ \\ \hline
			With SCF&-69.69& -68.12 &-33.98 &-267.32&-2.28\\
			Without SCF&-72.52&-56.18  &144.90 &-629.35&0.21
		\end{tabular}
	\label{Table_rot_power_molecule}
	\end{ruledtabular}
\end{table}
Table \ref{Table_rot_power_molecule} shows our computed gyration tensor (multiplied by the volume of the simulation cell $\Omega$), 
with and without SCF terms; as in crystals, the latter have a huge impact on some components. 
We also report the optical rotatory parameter $\beta$, which in molecular systems relates to
the rotatory power $\alpha(\omega)$ via \cite{GRIMME2001380,Rot_Autschbach}
\begin{equation}
\alpha(\omega)=\frac{N_A\omega^2}{Mc^2}\beta,\qquad \beta=\frac{\Omega}{4\pi}\frac{1}{2}\sum_{a}\frac{1}{3}g_{aa}.
\end{equation}
Here $N_A$ is the Avogadro number and
$M$ is the molar mass of the molecule.
Our computed value of $\beta$ almost exactly matches the value of $\beta=-2.29$ that was reported in Ref.~\cite{doi:10.1063/1.4865229}.
Although such a level of agreement gives us confidence in the correctness of our implementation, it 
may be to some extent coincidental, given the differences in our respective approximations and 
computational schemes.

In summary, we have presented a formulation of optical dispersion within the framework of density-functional
perturbation theory. Our methodology brings the first-principles calculation of the gyration tensor to the same level 
of accuracy and computational ease as standard linear-response properties, e.g., the dielectric tensor.
We have also discussed some formal aspects of the theory, e.g., the non-uniqueness of Eq. (\ref{Eq_E_gamma_EE}), which we relate to the gauge 
freedom of electromagnetism. 
As an outlook, a natural step forward consists in generalizing our method to 
finite frequencies, and to magnetic materials with broken time-reversal symmetry;
progress along these lines will be presented in a forthcoming publication.\\

\begin{acknowledgments}
We acknowledge support from Ministerio de Ciencia
e Innovaci\'on (MICINN-Spain) through
Grant No. PID2019-108573GB-C22;
from Severo Ochoa FUNFUTURE center of excellence (CEX2019-000917-S);
from Generalitat de Catalunya (Grant No. 2021 SGR 01519); and from
the European Research Council (ERC) under the European Union's
Horizon 2020 research and innovation program (Grant
Agreement No. 724529).
\end{acknowledgments}

\clearpage

\beginsupplement
\onecolumngrid
´
\section{Supplementary notes for ``Natural optical activity from
	density-functional perturbation theory''}

\begin{center}
	Asier Zabalo$^1$ and Massimiliano Stengel$^{1,2}$\\\vspace*{0.1cm}
	$^1$\textit{Institut de Ci\`encia de Materials de Barcelona (ICMAB-CSIC), Campus UAB, 08193 Bellaterra, Spain}\\
    $^2$\textit{ICREA-Instituci\'o Catalana de Recerca i Estudis Avan\c{c}ats, 08010 Barcelona, Spain}
\end{center}

\section{I. Computational details}
Our numerical results of the main text are obtained using the DFT
and DFPT implementations of the {\sc abinit} package with
the Perdew-Wang \cite{PhysRevB.45.13244} parametrization of the LDA. 
We use norm-conserving pseudopotentials from the Pseudo Dojo \cite{VANSETTEN201839} website and we regenerate them without exchange-correlation nonlinear core corrections using the ONCVPSP \cite{PhysRevB.88.085117} software. Spin-orbit coupling (SOC) is neglected in our first-principles calculations.
All the materials studied in this work, trigonal Se and $\alpha$-HgS and $\alpha$-SiO$_2$, belong to the point group 32, but they may crystalize in two enantiomorphic structures with space groups $P3_121$ and $P3_221$. These structures with opposite handedness
have the same rotatory power in magnitude, but with opposite sign.
In this work, we shall consider the $P3_121$ space group structure for the three crystals under study. The crystal structure is either set to the experimental one
(Ref. \cite{doi:10.1063/1.1743647} for Se), or relaxed to mechanical equilibrium 
until the forces are smaller than $10^{-6}$ Ha/bohr.
We use a plane-wave cutoff of 50 Ha for Se and $\alpha$-SiO$_2$ and 40 Ha for 
$\alpha$-HgS.
We find that the results are remarkably sensitive to this choice,
and on the exchange-correlation functional that is used in the
structural relaxation; details are provided in the next Section.

Regarding C$_4$H$_4$O$_2$, a box with sides of $a=35$ bohr ($\Omega=a^3$) is used in order to simulate an isolated molecule, with a plane-wave energy cutoff of 50 Ha. The non-relaxed coordinates are taken from Ref. \cite{doi:10.1063/1.4893991}. 
\begin{table}[h!]\label{Table_coordinates}
	\begin{ruledtabular}
		\caption{Cartesian coordinates (in bohr) of the C$_4$H$_4$O$_2$ molecule, from Ref. \cite{doi:10.1063/1.4893991}.}
		\begin{tabular}{crrr}
			C$_4$H$_4$O$_2$& \multicolumn{1}{c}{x} & \multicolumn{1}{c}{y} & \multicolumn{1}{c}{z}\\ \hline
			O$_1$& 0.643397 &1.236357 & -2.077495\\
			O$_2$&-0.643397&-1.236357 &-2.077495 \\
			C$_1$&-0.120062& 1.373532 &2.331775\\
			C$_2$&0.120062& 2.555671 & 0.116615\\
			C$_3$&0.120062& -1.373532 & 2.331775\\
			C$_4$&-0.120062& -2.555671 & 0.116615\\
			H$_1$&0.117087&4.574339  & -0.221634\\
			H$_2$&-0.496036& 2.427108 & 4.040414\\
			H$_3$&-0.117087&-4.574339  &-0.221634 \\
			H$_4$&0.496036& -2.427108 &4.040414 \\
		\end{tabular}
	\end{ruledtabular}
\end{table}
\section{II. Effect of the structural parameters and exchange and correlation functionals (LDA \lowercase{vs} GGA)}
\begin{figure}[h!]
	\includegraphics[width=0.49\linewidth]{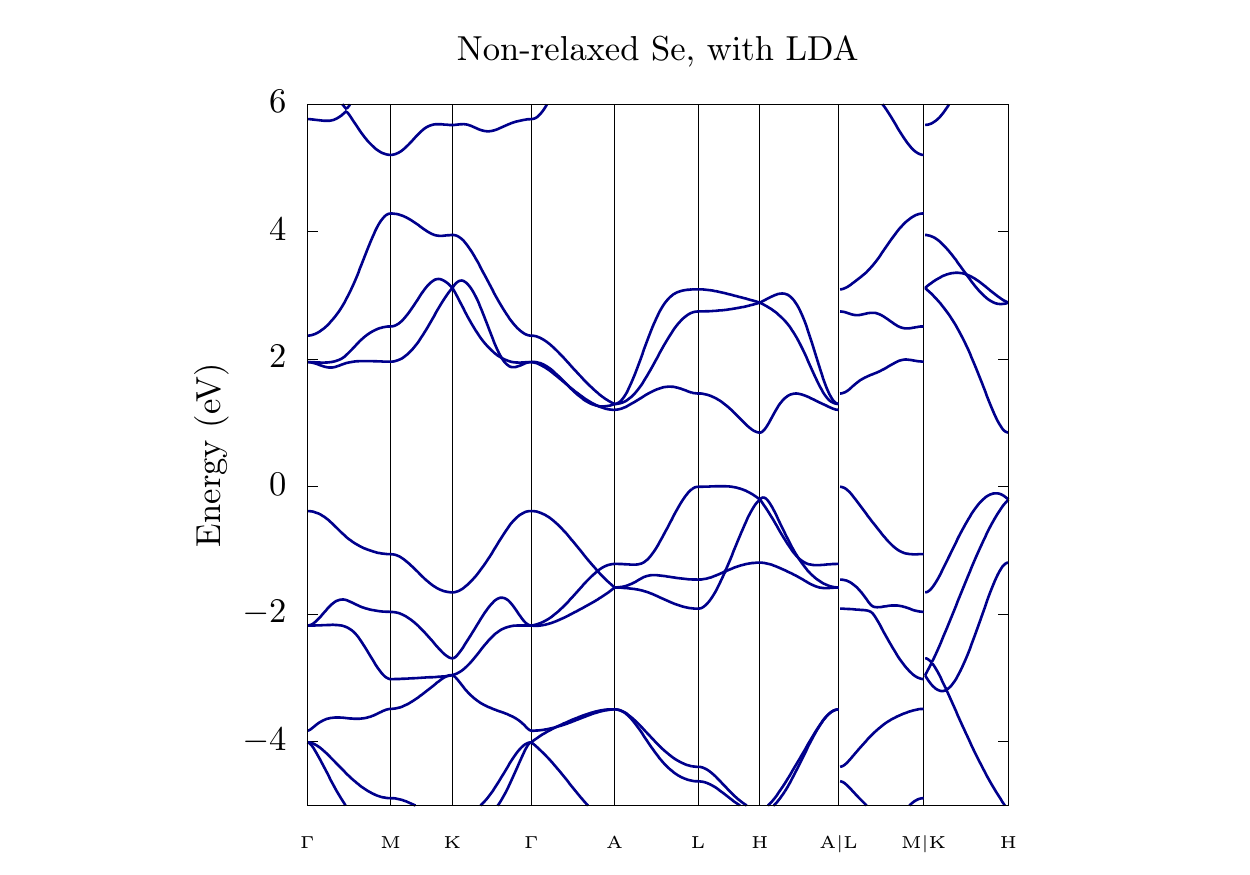}
	\includegraphics[width=0.49\linewidth]{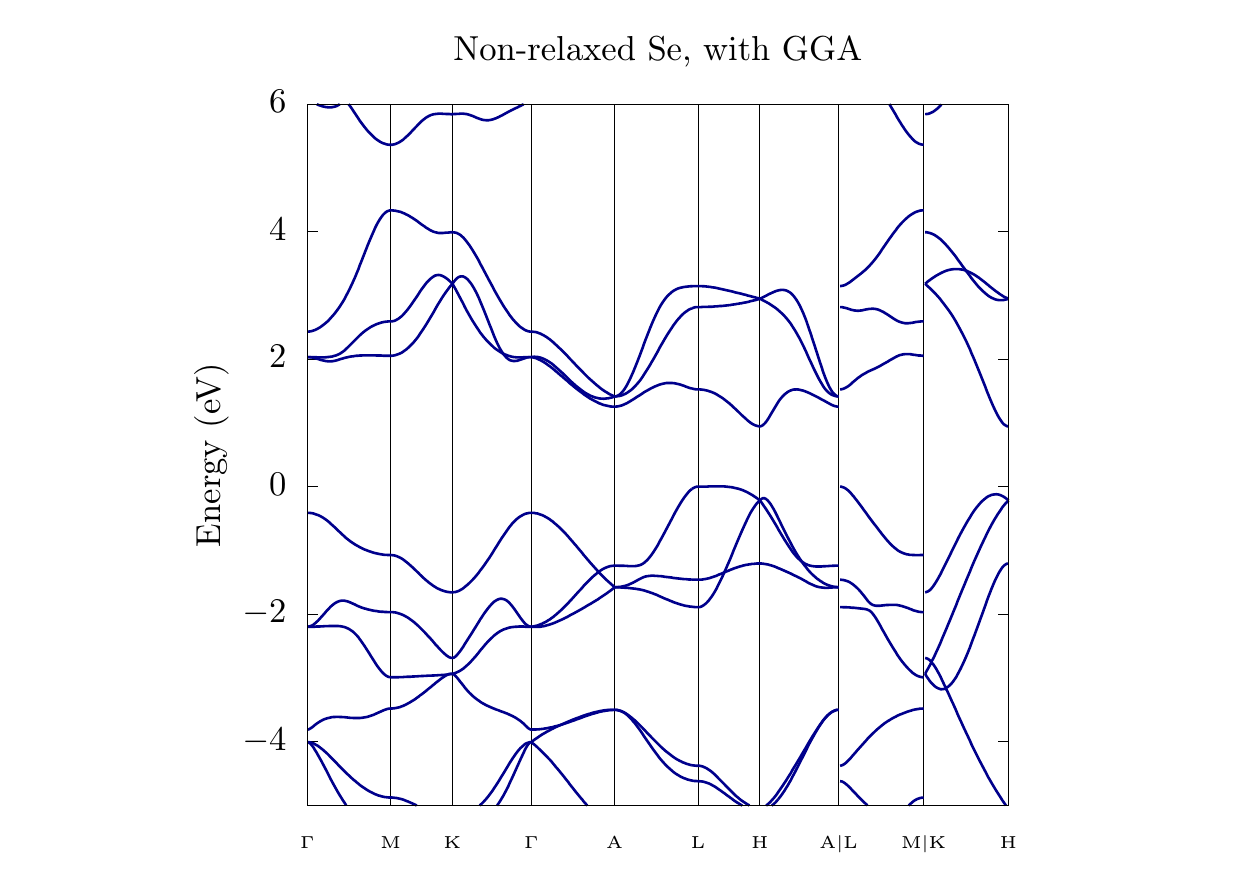}\\
	\includegraphics[width=0.49\linewidth]{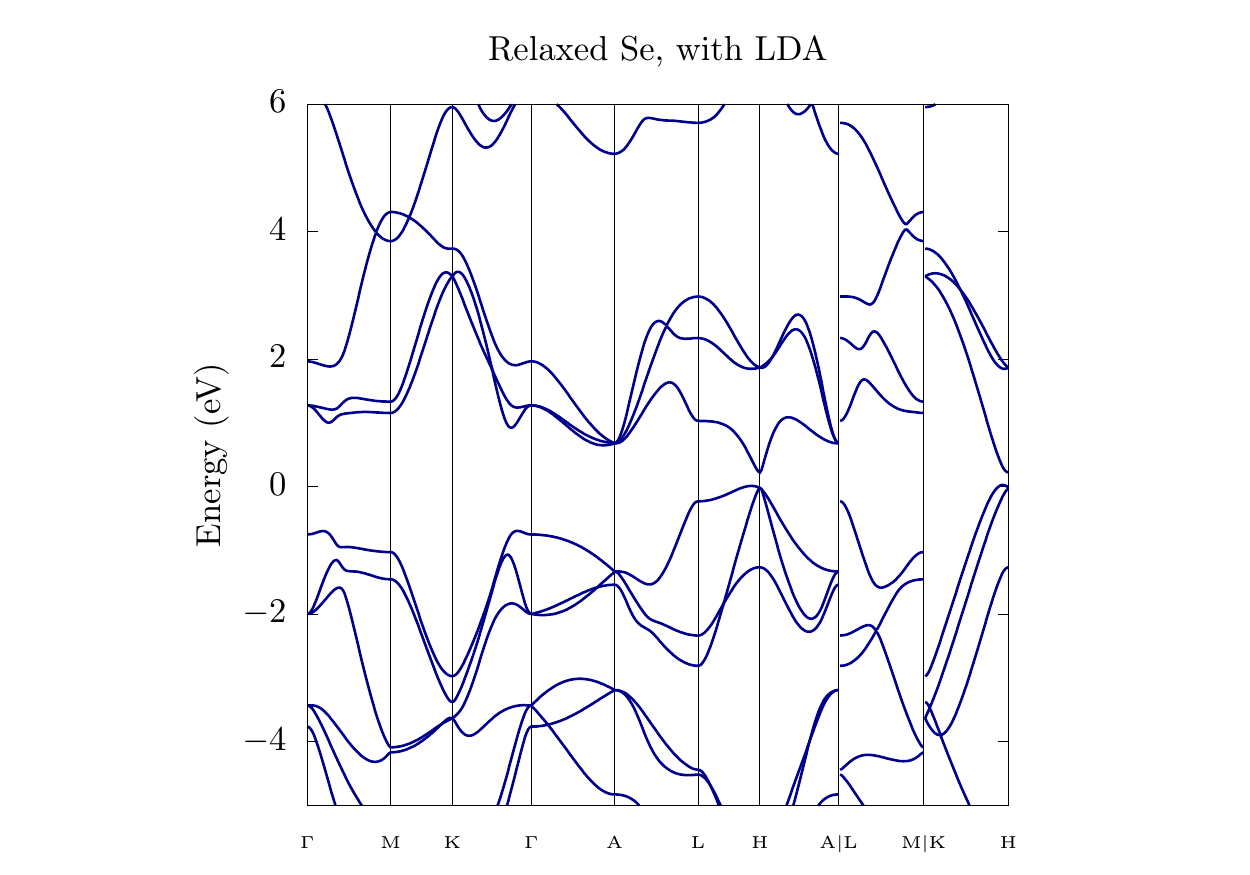}
	\includegraphics[width=0.49\linewidth]{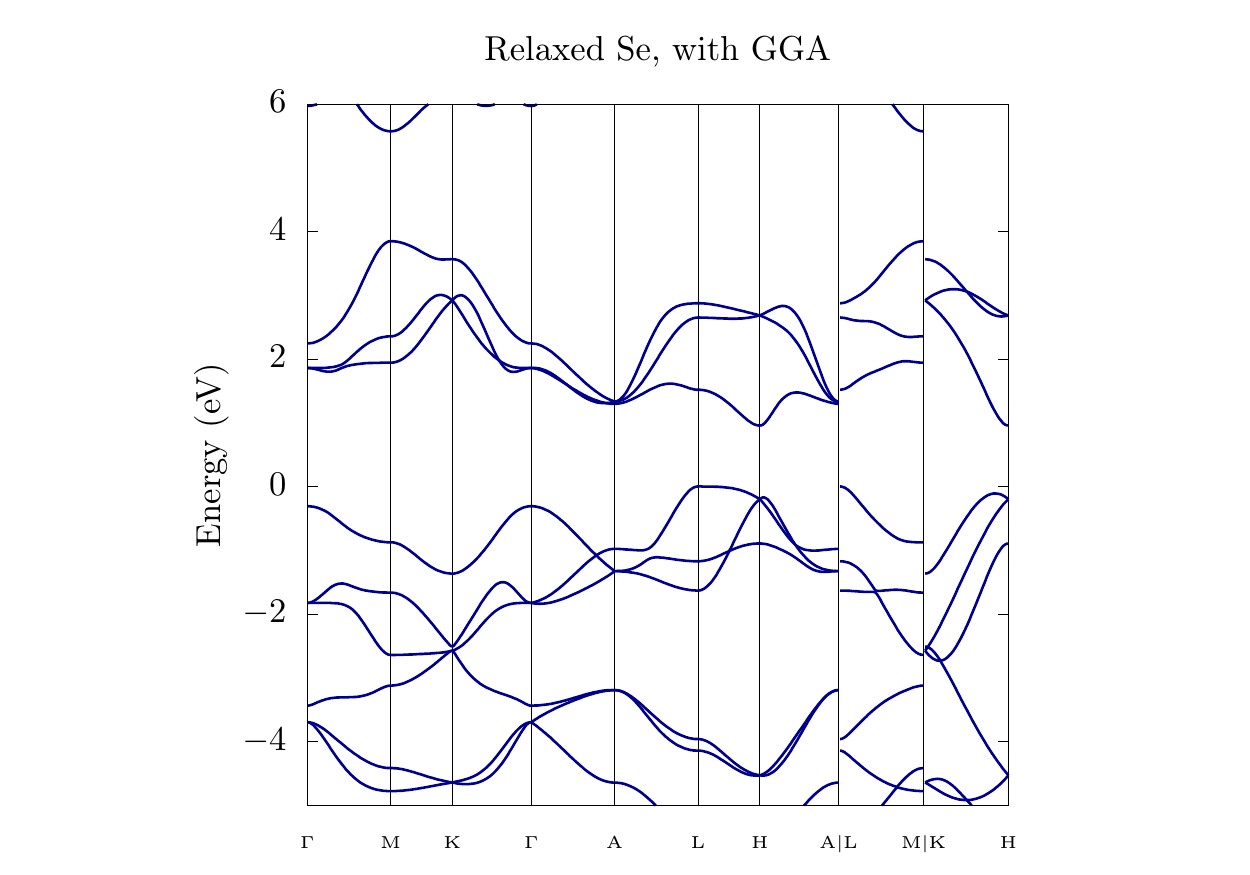}\\
	\caption{Band structure of trigonal Se, computed with LDA and GGA.}
	\label{Fig_BS_Se}
\end{figure}

\begin{figure}[t!]
	\includegraphics[width=0.49\linewidth]{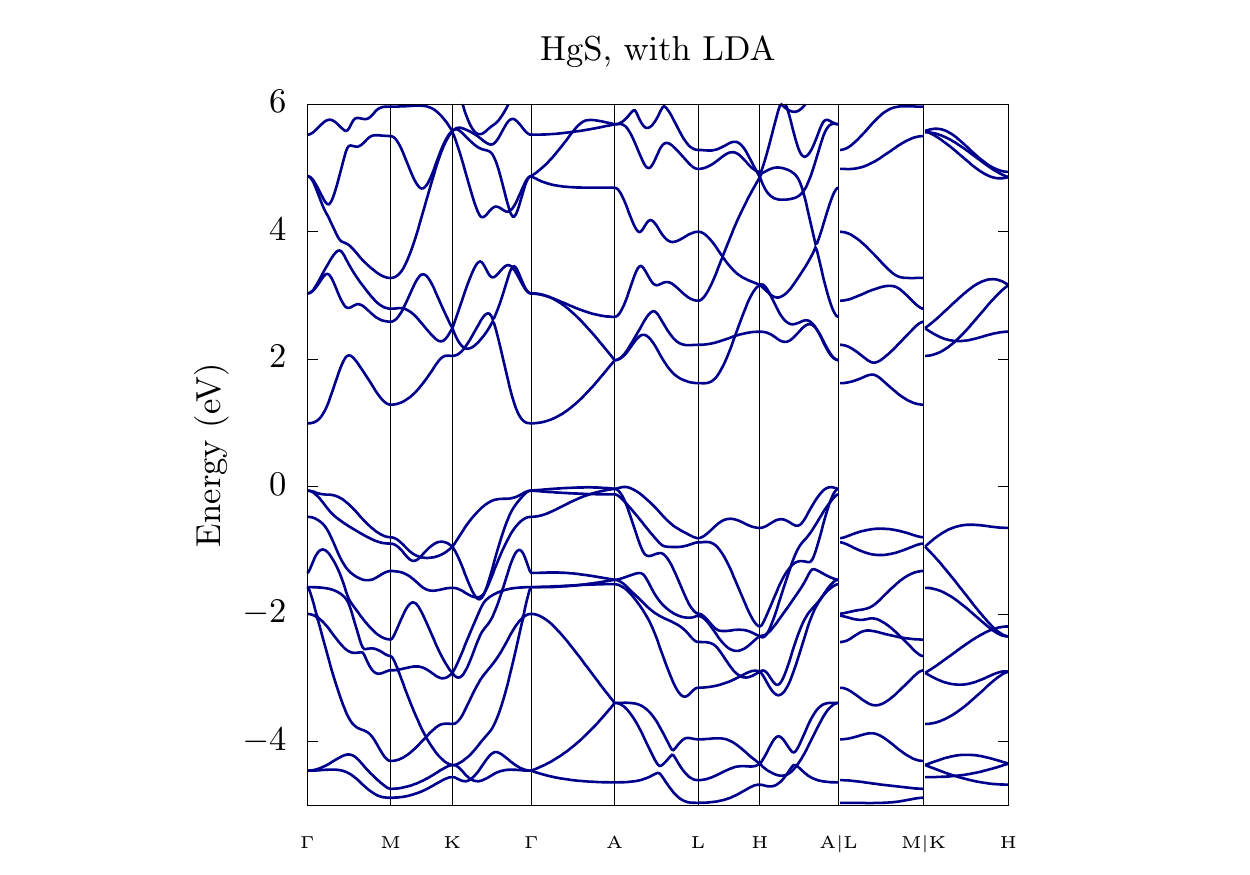}
	\includegraphics[width=0.49\linewidth]{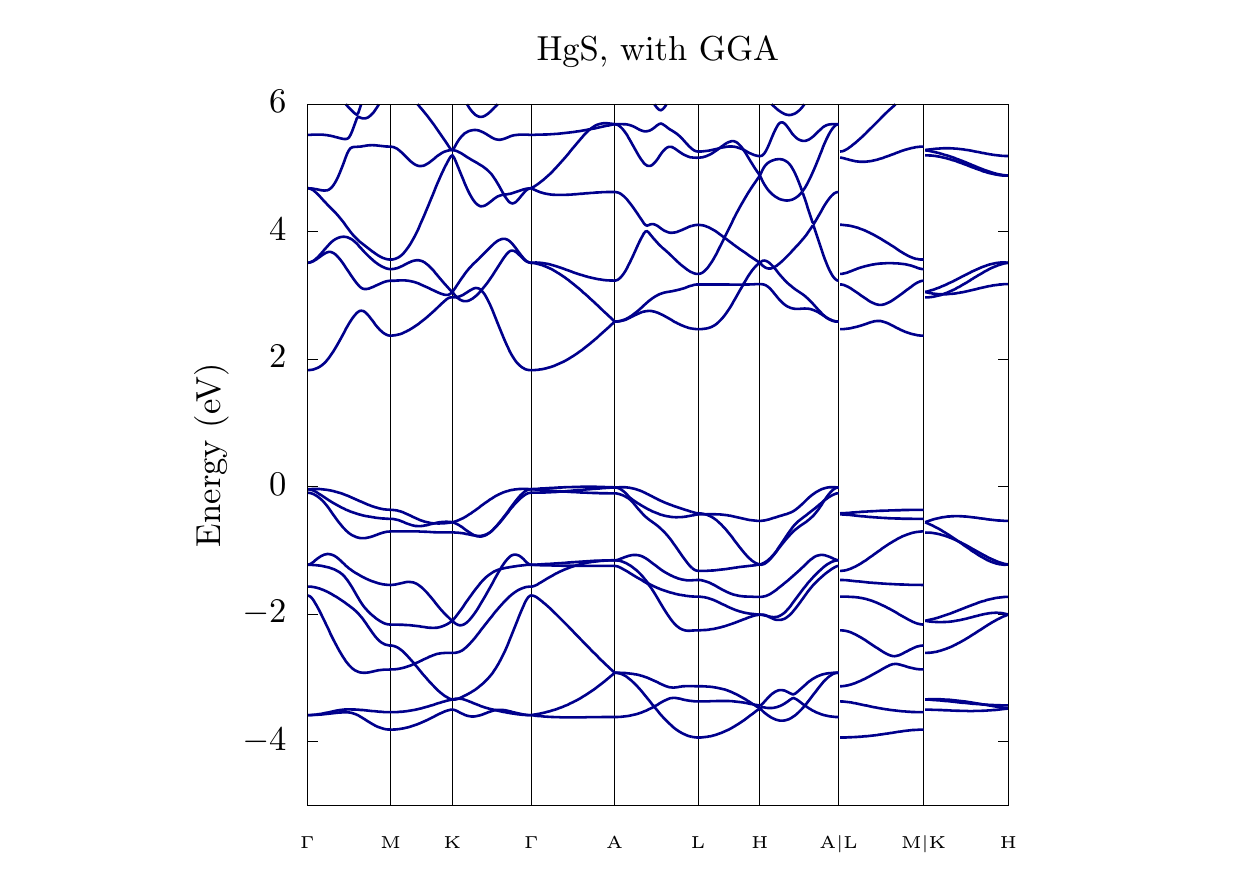}\\
	\caption{Band structure of $\alpha$-HgS, computed with LDA and GGA.}
	\label{Fig_BS_HgS}
\end{figure}

\begin{figure}[t!]
	\includegraphics[width=0.49\linewidth]{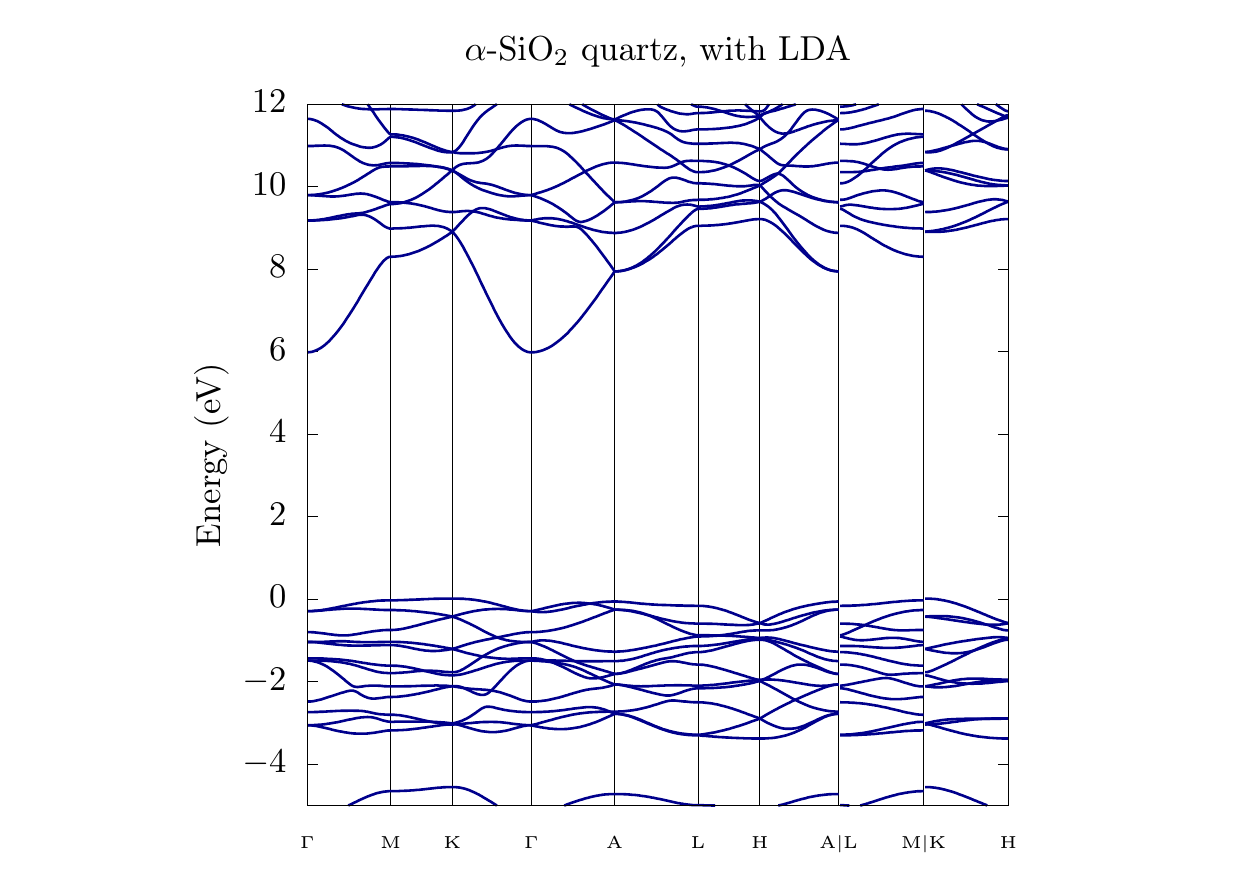}
	\includegraphics[width=0.49\linewidth]{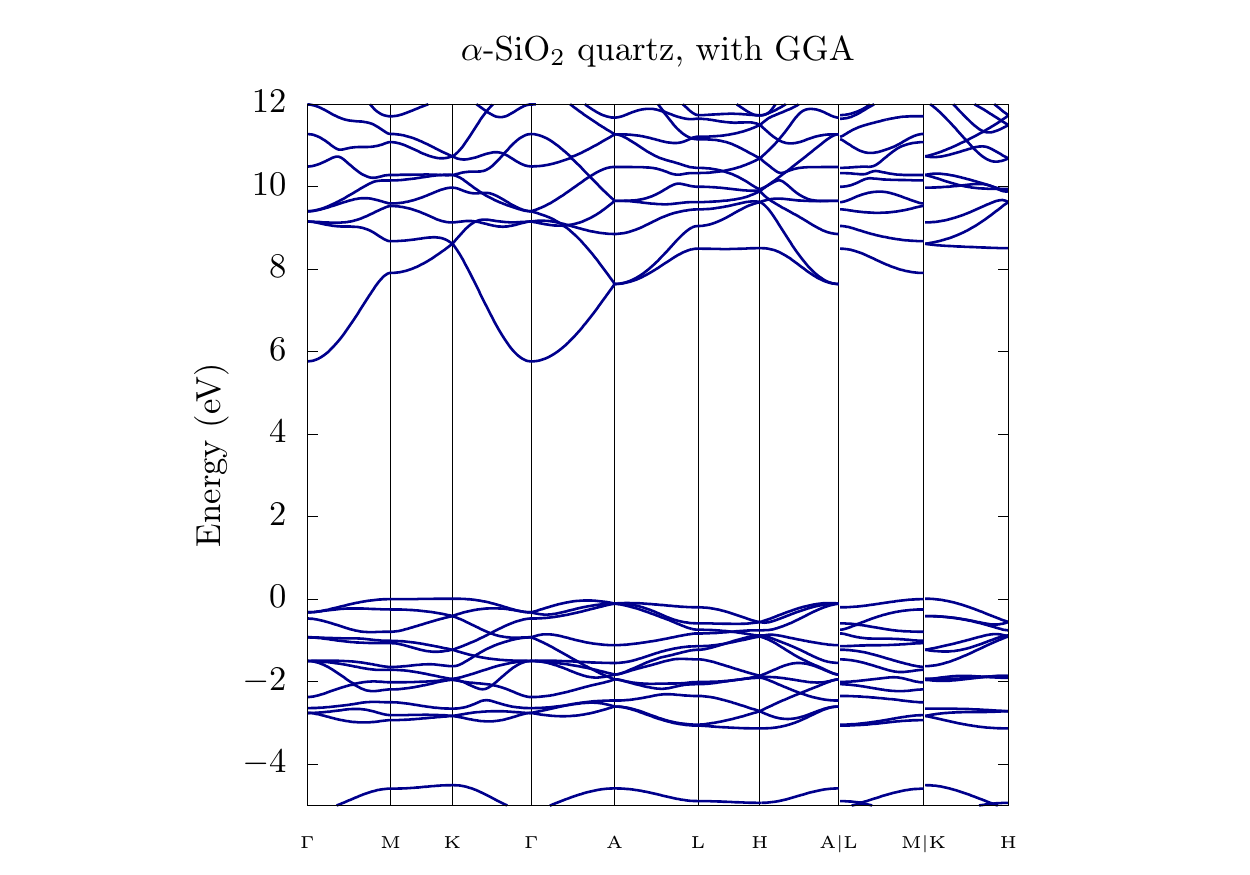}\\
	\caption{Band structure of $\alpha$-SiO$_2$, computed with LDA and GGA.}
	\label{Fig_BS_quartz}
\end{figure}

The aim of this section is to examine how the exchange and correlation functional affects the structural characteristics of the solids being analyzed. We shall see how going from LDA to GGA changes the structural parameters and the electronic band structure of Se, $\alpha$-HgS and $\alpha$-SiO$_2$ and see how this affects the dielectric tensor, and hence, the natural optical activity. Table \ref{Table_GGA_LDA} shows the structural parameters for different cases. As expected, GGA gives a larger
unit cell when the forces are relaxed and, in general, a larger electronic band gap (the only exception here is $\alpha$-SiO$_2$).
\begin{table}[h!]\label{Table_GGA_LDA}
	\begin{ruledtabular}
		\caption{Structural parameters and electronic band gap of Se, $\alpha$-HgS and $\alpha$-SiO$_2$, with LDA and GGA. In the Structure column,
			“exp” refers to the experimental structure, while Se (LDA)
			means that the structure was relaxed with LDA, for example.
			The Functional column specifies the exchange and correlation
			functional used for computing the electronic band structure.}
		\begin{tabular}{ccrrrr}
			Structure& Functional&$a$ (bohr) & $c$ (bohr) & $\Omega$ (bohr$^3$)& band gap (eV) \\ \hline
			Se (exp)&LDA& 8.201 &9.354 & 544.894&0.850\\
			Se (exp)&GGA& 8.201 &9.354 & 544.894&0.948\\
			Se (LDA) &LDA& 7.431  & 9.695 &463.605&0.225 \\
			Se (GGA)&GGA&8.463   & 9.564 & 593.187&0.961 \\ \hline
			$\alpha$-HgS (LDA)&LDA& 7.629& 17.530&883.498 &0.995 \\
			$\alpha$-HgS (GGA)&GGA&8.434 &18.430 &1135.414 &1.829\\ \hline
			$\alpha$-SiO$_2$ (LDA)&LDA& 9.174 & 10.110& 736.974&5.987  \\
			$\alpha$-SiO$_2$ (GGA)&GGA& 9.489 &10.411 & 811.792& 5.768\\
		\end{tabular}
	\end{ruledtabular}
\end{table}
Fig. \ref{Fig_BS_Se}, \ref{Fig_BS_HgS} and \ref{Fig_BS_quartz} show the electronic band structure 
of Se, $\alpha$-HgS and $\alpha$-SiO$_2$, respectively. For Se, we study four cases: the non-relaxed structure and the relaxed structure, both with LDA and GGA. When the structure is relaxed with LDA, the band gap drastically decreases and, in fact, it becomes so small (around 0.2 eV) that even standard DFPT calculations, e.g. the quadrupoles, fail to converge. In general, we observe that the larger the band gap, the
faster the convergence is, which might explain the slightly faster convergence of the natural optical activity with GGA, in contrast to LDA, that is shown in Sec. III. The most intriguing discovery arises when examining the impact of structural parameters on the dielectric tensor, as shown in Table S3. Surprisingly, the choice of the exchange and correlation functional has minimal effect on the independent components of the dielectric tensor. However, the structural parameters play a significant role here, particularly evident in the case of $\alpha$-HgS. Although the effect is less dramatic for Se and $\alpha$-SiO$_2$, it becomes more pronounced when calculating the natural optical activity tensor, which is just the first-order derivative of the dielectric tensor with respect to $\mathbf{q}$ (see Table II of main text).
\begin{table}[b!]\label{Table_Dielectric}
	\begin{ruledtabular}
		\caption{Comparison between LDA and GGA for the independent components of the dielectric permittivity tensor for Se, 
			$\alpha$-HgS and $\alpha$-SiO$_2$, for different structures, in Hartree atomic units. In the Structure column,
			“exp” refers to the experimental structure, while Se (LDA)
			means that the structure was relaxed with LDA, for example.}
		\begin{tabular}{ccrrr}
			Structure&\multicolumn{2}{c}{$\epsilon_{11}$} & \multicolumn{2}{c}{$\epsilon_{33}$} \\ 
			&LDA&GGA&LDA&GGA\\ \hline
			Se (exp) & 9.05817 &8.44465 & 14.66883& 13.89905\\
			Se (GGA) & 9.19622 & 8.53039&14.77391 & 13.92631\\ \hline
			$\alpha$-HgS (LDA) & 10.07620 & 9.40749& 12.10559&11.23652 \\
			$\alpha$-HgS (GGA) & 5.89176 & 5.56827& 8.21707& 7.72728\\ \hline
			$\alpha$-SiO$_2$ (LDA) & 2.48932 &2.45577 & 2.52075&2.48670\\
			$\alpha$-SiO$_2$ (GGA) &2.42803 & 2.39744& 2.45630& 2.42505\\ \hline
		\end{tabular}
	\end{ruledtabular}
\end{table}
To gain further insight into the effect of the structural parameters, in Table
S4 we show the individual contributions to the NOA tensor for different structures. In general, the effect of the structure is clearly noticeable in the individual terms, which can be easily appreciated for $\alpha$-HgS. However, even if the individual contributions differ
to a large degree (see the $\mathcal{Y}$ and $\mathcal{W}$ terms in Se), an accidental
cancellation of several terms can lead to a similar final result for the NOA tensor components.
\begin{table}[h!]\label{Table_T_terms}
	\begin{ruledtabular}
		\caption{Contribution to the $g_{33}$ tensor component (in bohr) from different terms of Eq. (10) and Eq. (11) of the main text. We are showing the results obtained with the LDA exchange and correlation functional for the computation of the NOA, in all cases.}
		\begin{tabular}{crrrrr}
			Structure&\multicolumn{1}{c}{$E_\text{elst}$} & \multicolumn{1}{c}{$\mathcal{X}$} &
			\multicolumn{1}{c}{$\mathcal{Y}$} & 
			\multicolumn{1}{c}{$\mathcal{W}$}& \multicolumn{1}{c}{$g_{33}$}\\ \hline
			Se (exp)& -0.164 & -0.836 & -1.884 & 0.973&-1.910\\
			Se (GGA)& -0.169 & -0.729 & -2.482& 1.578& -1.802\\\hline
			$\alpha$-HgS (LDA) &-0.095 & -1.106 &-0.801 & 0.149&  -1.861 \\
			$\alpha$-HgS (GGA) & -0.204&-0.054&-0.479& 0.672& -0.065\\\hline
			$\alpha$-SiO$_2$ (LDA)&0.010  & 0.020 & 0.099  & -0.005&0.125\\
			$\alpha$-SiO$_2$ (GGA)&0.016& 0.029 &0.118  & 0.006&0.168
		\end{tabular}
	\end{ruledtabular}
\end{table}
\clearpage
\section{III. Convergence study of the natural optical activity tensor}
As a further step for the validation of our method, we plot
the independent components of the gyration tensor, $g_{11}$
and $g_{33}$, for trigonal Se, $\alpha$-HgS and $\alpha$-SiO$_2$ . Fig. S4, S5 and S6 show
the obtained numerical results as a function of the plane-wave cutoff energy and the $\mathbf{k}$ points mesh resolution. For comparison purposes, we also compute the natural optical activity
with the PBE \cite{PhysRevLett.77.3865} parametrization of the GGA. (For trigonal Se, the experimental structure is used both for LDA and GGA; whereas for $\alpha$-HgS and $\alpha$-SiO$_2$, the structure is relaxed until the forces 
are smaller than $10^{-6}$ Ha/bohr, either with LDA or GGA.)
\subsection{Trigonal Se}
\begin{figure}[h!]
	\includegraphics[width=0.6\linewidth]{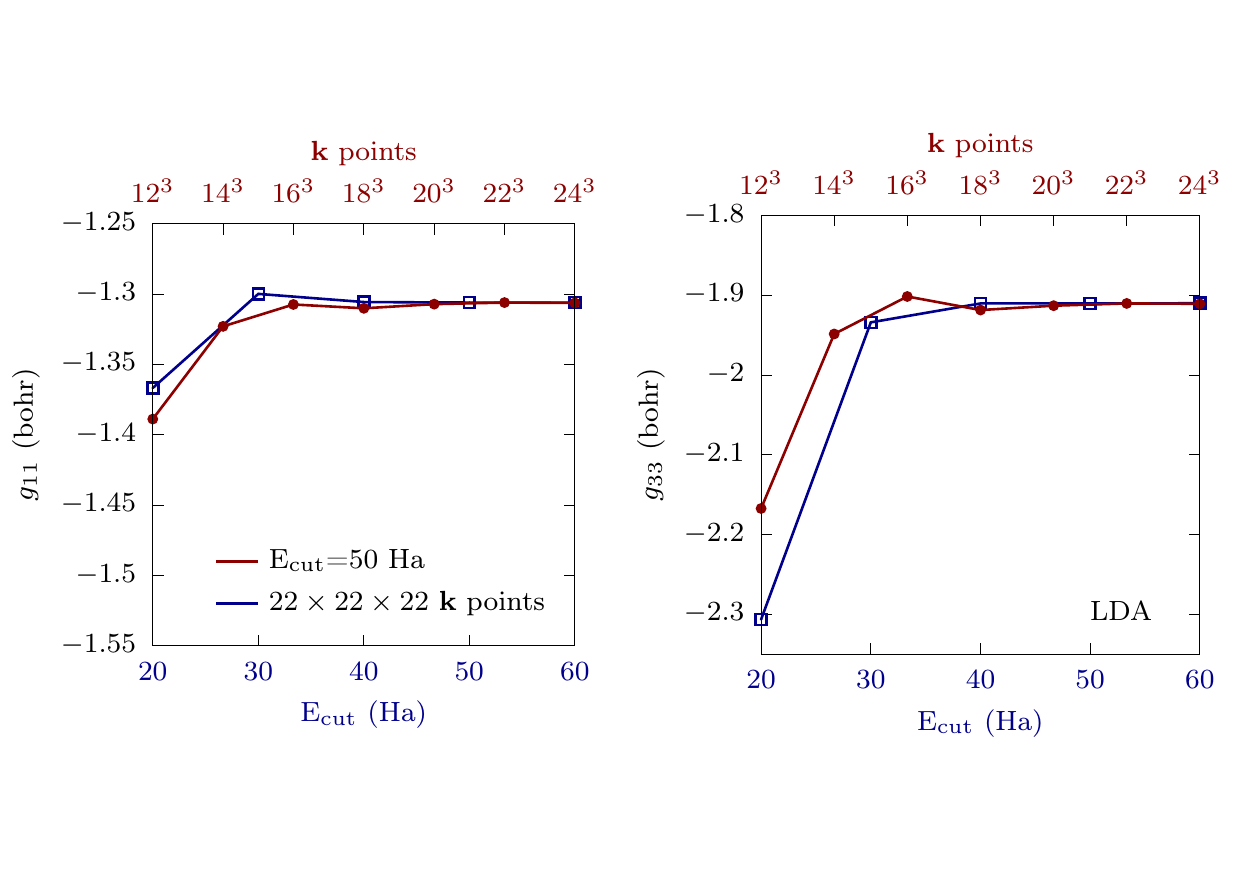}\\
	\includegraphics[width=0.6\linewidth]{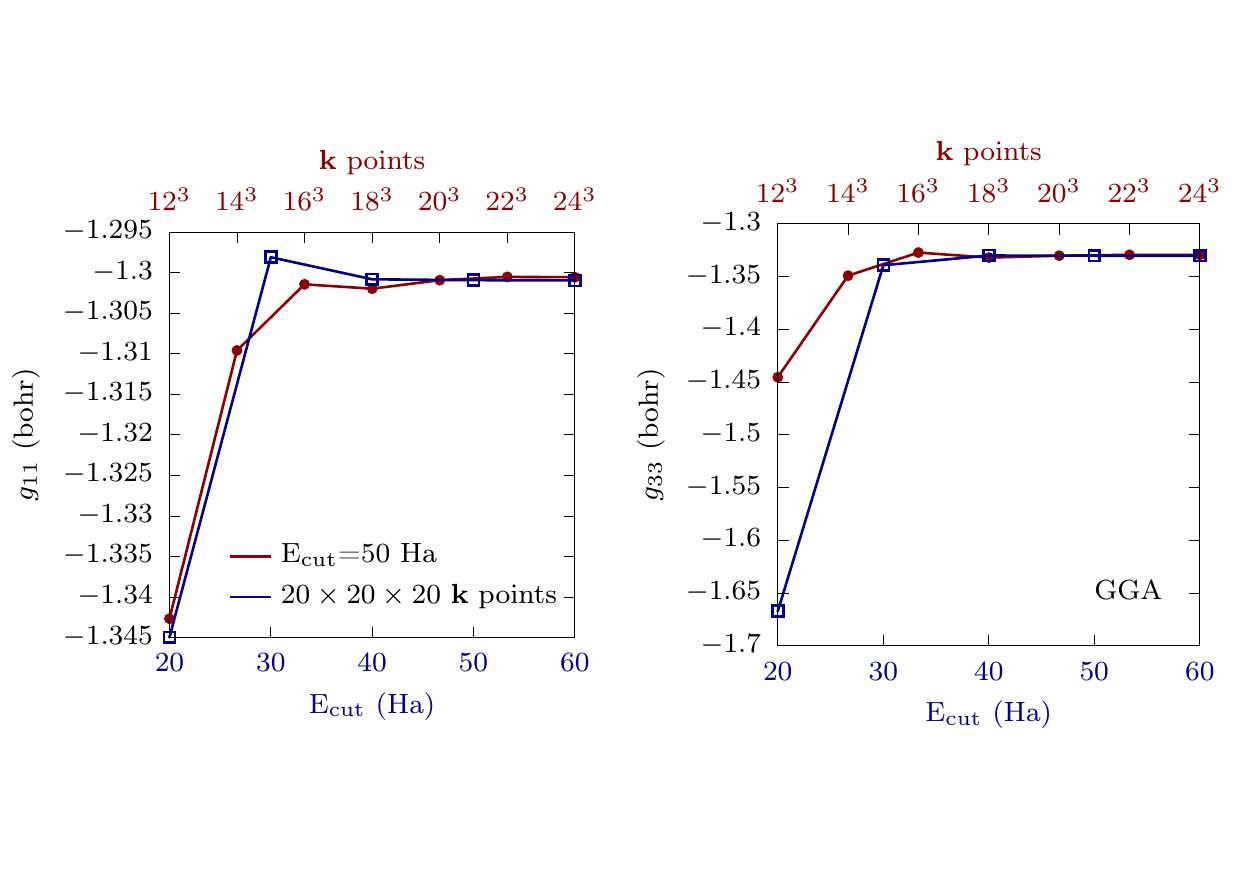}\\
	\caption{Convergence of the independent components of the gyration tensor, $g_{11}$ and $g_{33}$, of non-relaxed trigonal Se, with respect to the plane-wave cutoff and the density of the $\mathbf{k}$-point mesh. The top panel shows the obtained results with LDA, whereas the bottom panel shows the results obtained with GGA.}
	\label{Fig_convergence_Se}
\end{figure}
\clearpage
\subsection{$\alpha$-HgS}
\begin{figure}[h!]
	\includegraphics[width=0.6\linewidth]{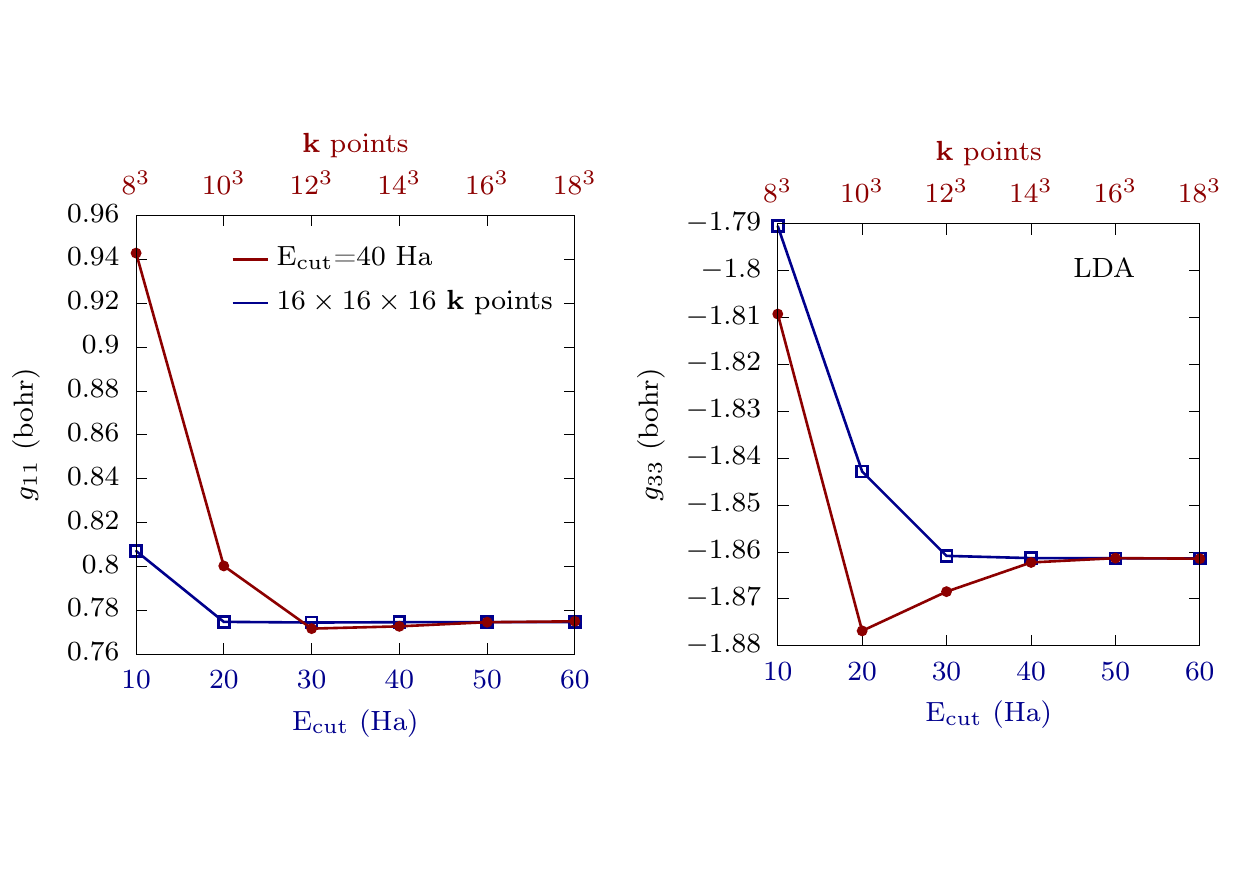}\\
	\includegraphics[width=0.6\linewidth]{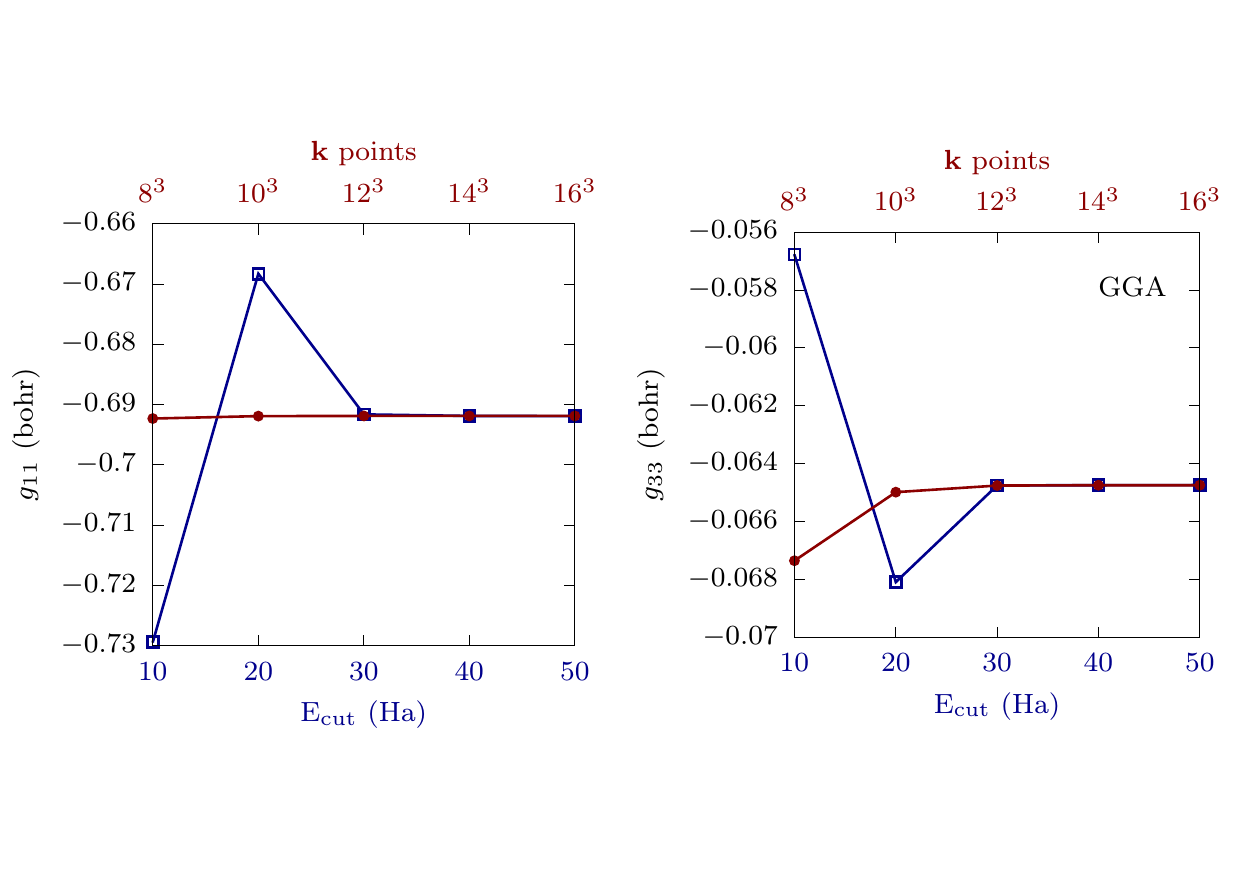}\\
	\caption{Convergence of the independent components of the gyration tensor, $g_{11}$ and $g_{33}$, of $\alpha$-HgS, with respect to the plane-wave cutoff and the density of the $\mathbf{k}$-point mesh. The top panel shows the obtained results with LDA, whereas the bottom panel shows the results obtained with GGA.}
	\label{Fig_convergence_HgS}
\end{figure}
\clearpage
\subsection{$\alpha$-SiO$_2$}
\begin{figure}[h!]
	\includegraphics[width=0.6\linewidth]{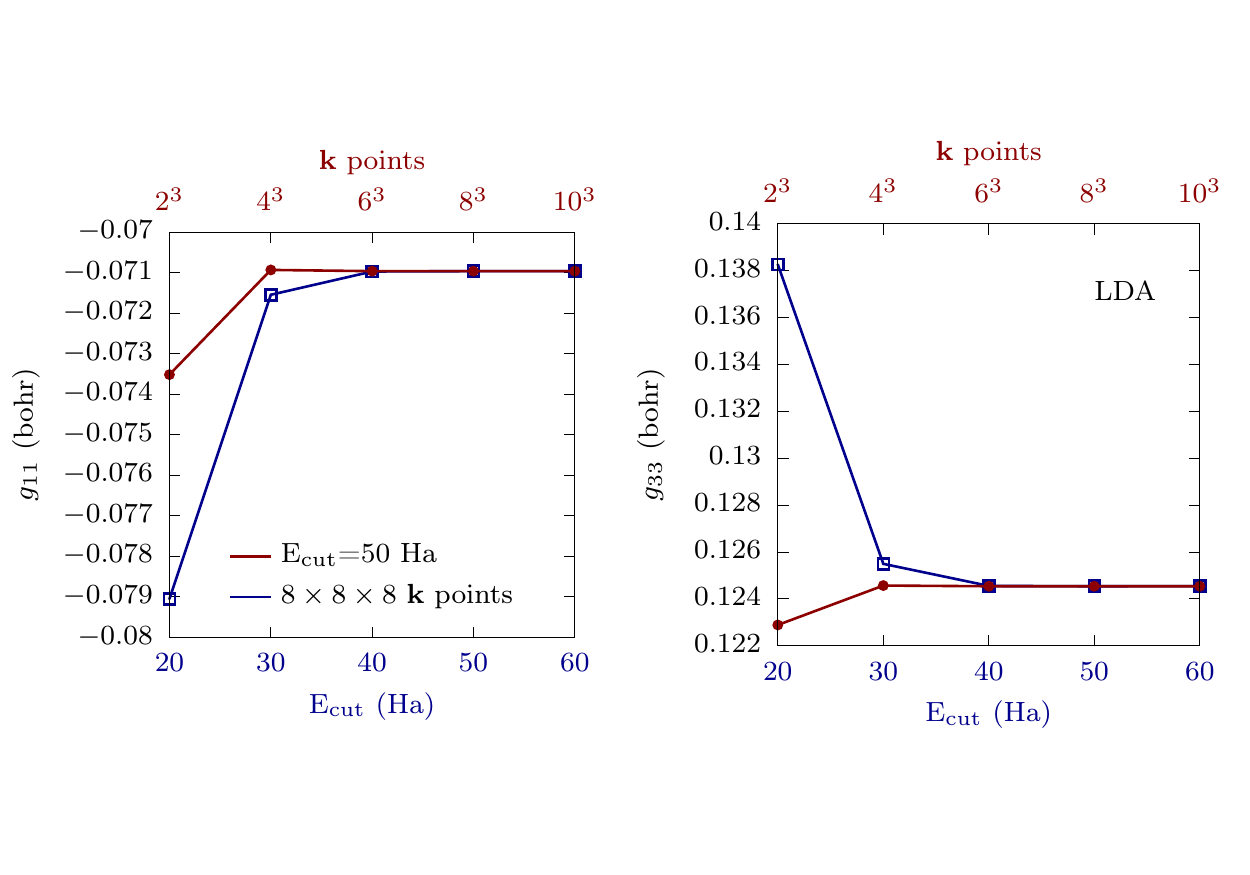}\\
	\includegraphics[width=0.6\linewidth]{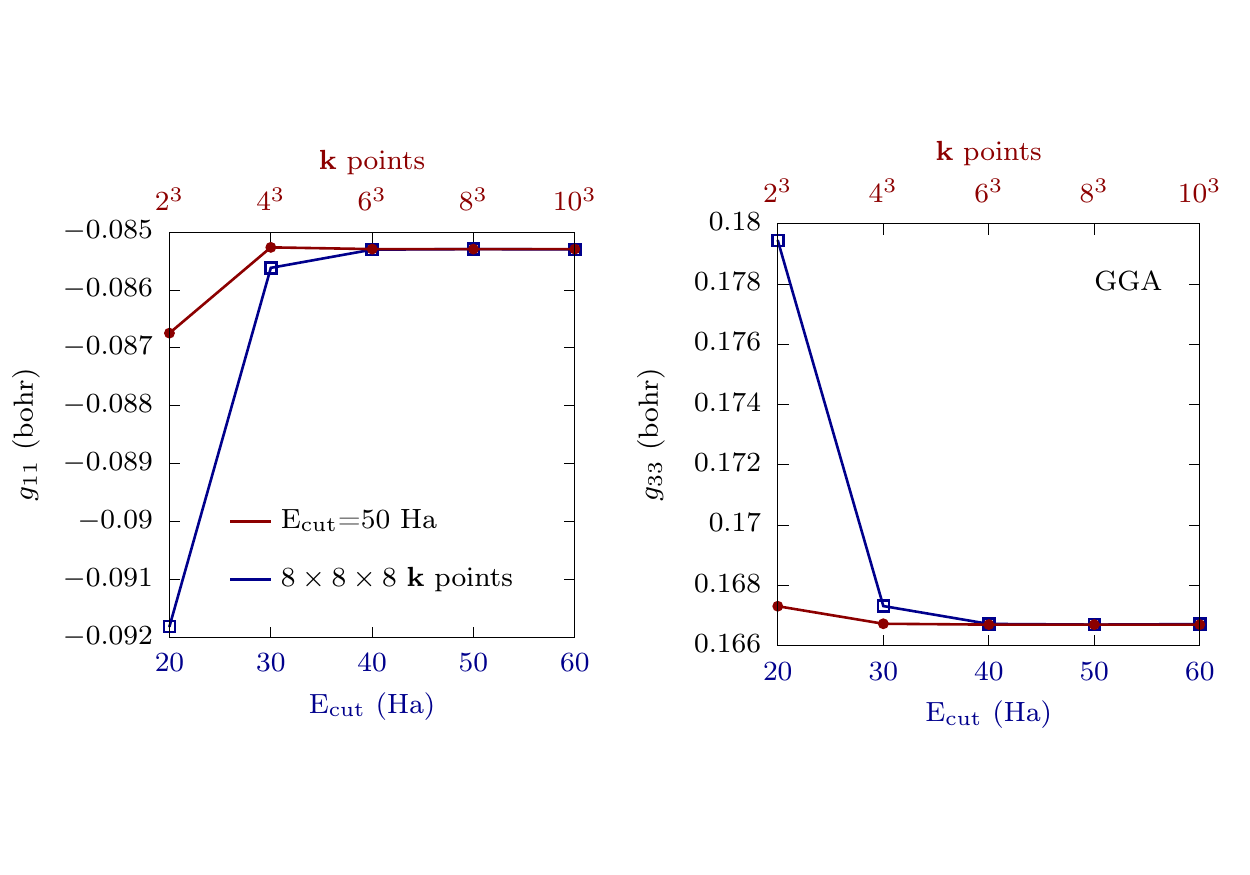}\\
	\caption{Convergence of the independent components of the gyration tensor, $g_{11}$ and $g_{33}$, of $\alpha$-SiO$_2$, with respect to the plane-wave cutoff and the density of the $\mathbf{k}$-point mesh. The top panel shows the obtained results with LDA, whereas the bottom panel shows the results obtained with GGA.}
	\label{Fig_convergence_quartz}
\end{figure}
\clearpage
\section{IV. Response to a long-wavelength magnetic vector potential field}\label{Appendix_vector_potential}
The wave-function response to a long-wavelength magnetic vector potential field
is given by the following Sternheimer equation \cite{PhysRevX.9.021050,PhysRevB.105.094305}
\begin{equation}\label{Eq_vetor_pot}
(\hat{H}_\mathbf{k}^{(0)}+a\hat{P}_\mathbf{k}-\epsilon_\mathbf{k}^{(0)})
\ket*{u_{m\mathbf{k},\gamma}^{A_\beta}}=
-\hat{Q}_\mathbf{k}\hat{O}_\mathbf{k}^{\alpha\gamma}\ket*{u_{m\mathbf{k}}^{(0)}},
\end{equation}
where $\hat{H}^{(0)}_\mathbf{k}$ is the ground state Hamiltonian, 
$\epsilon_{m\mathbf{k}}^{(0)}$ is an unperturbed energy eigenvalue and 
$\hat{Q}_\mathbf{k}=1-\hat{P}_\mathbf{k}$ is the conduction-band projector with
$\hat{P}_\mathbf{k}=\sum_n \ket*{u_{n\mathbf{k}}^{(0)}}\bra*{u_{n\mathbf{k}}^{(0)}}$. 
In Eq. (\ref{Eq_vetor_pot}), $a$ is a parameter with the dimension of energy
that ensures stability \cite{PhysRevX.9.021050}. The perturbing operator 
$\hat{O}_\mathbf{k}^{\alpha\gamma}$ is given by 
\begin{equation}\label{Eq_O}
\hat{O}_\mathbf{k}^{\alpha\gamma}=\partial_\gamma \hat{H}_\mathbf{k}^{(0)}
\partial_\alpha\hat{P}_\mathbf{k}-
\partial_\gamma\hat{P}_\mathbf{k}\partial_\alpha\hat{H}_\mathbf{k}^{(0)}
+\frac{1}{2}\partial^2_{\alpha\gamma}\hat{H}_\mathbf{k}^{(0)},
\end{equation}
where $\partial^2_{\gamma\alpha}\equiv \partial^2/\partial_{k_\alpha}\partial_{k_\gamma}$ is a second derivative in $\mathbf{k}$ space. (Notice that  Eq. (\ref{Eq_O}) gives the correct response to a long-wavelength magnetic vector potential field. There is a typo in Ref. \cite{PhysRevB.105.094305}, which makes Eq. (\ref{Eq_O}) differ from the expression given in Ref. \cite{PhysRevB.105.094305} by a half factor.)

The perturbing operator $\hat{O}_\mathbf{k}^{\beta\gamma}$ can be expressed as a sum of a symmetric $\hat{O}_\mathbf{k}^{S,\beta\gamma}$ and an antisymmetric 
$\hat{O}_\mathbf{k}^{A,\beta\gamma}$ part with respect to $\beta\leftrightarrow\gamma$ exchange, in a way that the symmetric part recovers 
the usual $d^2/dkdk$ wave functions, 
$\ket*{\partial^2_{\beta\gamma}u_{m\mathbf{k}}^{(0)}}$,
and the antisymmetric part recovers
the wave function response to a uniform orbital $\mathbf{B}$ field,
$\ket*{u_{m\mathbf{k}}^{B_\delta}}$, such that
\begin{equation}
\hat{O}_\mathbf{k}^{\beta\gamma}=\frac{1}{2}\big(
\hat{O}_\mathbf{k}^{S,\beta\gamma}+\hat{O}_\mathbf{k}^{A,\beta\gamma}
\big),
\end{equation}
where $\hat{O}_\mathbf{k}^{S,\beta\gamma}$ is the perturbing operator that enters the
Sternheimer equation for the $\ket*{\partial^2_{\beta\gamma}u_{m\mathbf{k}}^{(0)}}$ wave functions,
\begin{equation}\label{Eq_S_operator}
\hat{O}_\mathbf{k}^{S,\beta\gamma}=
\partial_\gamma\hat{H}^{(0)}_\mathbf{k}\partial_\beta\hat{P}_\mathbf{k}
+\partial_\beta\hat{H}_\mathbf{k}^{(0)}\partial_\gamma\hat{P}_\mathbf{k}
-\partial_\gamma\hat{P}_\mathbf{k}\partial_\beta\hat{H}^{(0)}_\mathbf{k}
-\partial_\beta\hat{P}_\mathbf{k}\partial_\gamma\hat{H}^{(0)}_\mathbf{k}
+\partial^2_{\beta\gamma}\hat{H}^{(0)}_\mathbf{k},
\end{equation}
and $\hat{O}_\mathbf{k}^{A,\beta\gamma}$ is intimately related to the perturbing operator that
enters the Sternheimer
equation for the wave function response to a uniform orbital magnetic field, 
$\hat{O}_\mathbf{k}^{B_\delta}$, as defined by 
Essin \textit{et al.} [\onlinecite{PhysRevB.81.205104}],
\begin{equation}
\hat{O}_\mathbf{k}^{B_\delta}=\frac{i}{2}\epsilon_{\delta\beta\gamma}\hat{O}_\mathbf{k}^{A,\beta\gamma},
\end{equation}
where
\begin{equation}
\hat{O}_\mathbf{k}^{A,\beta\gamma}=\partial_\gamma\hat{H}_\mathbf{k}^{(0)}\partial_\beta\hat{P}_\mathbf{k}
-\partial_\beta\hat{H}_\mathbf{k}^{(0)}\partial_\gamma\hat{P}_\mathbf{k}
-\partial_\gamma\hat{P}_\mathbf{k}\partial_\beta\hat{H}_\mathbf{k}^{(0)}
+\partial_\beta\hat{P}_\mathbf{k}\partial_\gamma\hat{H}_\mathbf{k}^{(0)}.
\end{equation}
Using the above relations one can readily verify that Eq. (15), (16) and (17) of the main text are nicely fulfilled.

\section{V. Natural optical activity without SCF terms}
\label{Appendix_without_SCF}
We shall use tilded symbols to indicate that SCF terms are excluded. Without self-consistency, the natural optical activity tensor reads as
\begin{equation}\label{Eq_tilde_eta}
\tilde{\eta}_{\alpha\beta\gamma}=-\frac{4\pi}{\Omega}\text{Im}
\tilde{E}_\gamma^{\mathcal{E}_\alpha^*\mathcal{E}_\beta},
\end{equation}
where
\begin{equation}
\tilde{E}_\gamma^{\mathcal{E}_\alpha^*\mathcal{E}_\beta}=2s\int_\text{BZ}
[d^3k] \tilde{E}_{\mathbf{k},\gamma}^{\mathcal{E}_\alpha^*\mathcal{E}_\beta}.
\end{equation}
The wave function term is given by
\begin{equation}
\begin{split}
\tilde{E}_{\mathbf{k},\gamma}^{\mathcal{E}_\alpha^*\mathcal{E}_\beta}&= 
\tilde{\mathcal{X}}_\mathbf{k}^{\mathcal{E}_\alpha k_\gamma\mathcal{E}_\beta}
+\tilde{\mathcal{W}}_\mathbf{k}^{\alpha,\beta\gamma}
+\big[\tilde{\mathcal{W}}_\mathbf{k}^{\beta,\alpha\gamma}\big]^*
\\
&=\sum_m \Big(\bra*{\tilde{u}_{m\mathbf{k}}^{\mathcal{E}_\alpha}}\hat{H}_\mathbf{k}^{k_\gamma
}\ket*{\tilde{u}_{m\mathbf{k}}^{\mathcal{E}_\beta}}+
\bra*{\tilde{u}_{m\mathbf{k}}^{\mathcal{E}_\alpha}}\ket*{iu_{m\mathbf{k},\gamma}^{A_\beta}}+
\bra*{iu_{m\mathbf{k},\gamma}^{A_\alpha}}\ket*{\tilde{u}_{m\mathbf{k}}^{\mathcal{E}_\beta}}\Big).
\end{split}
\end{equation}
In general, when self-consistency is taken into account, the first-order wave
functions $\ket*{u_{m\mathbf{k}}^{\mathcal{E}_\alpha}}$ are obtained by solving the following Sternheimer equation,
\begin{equation}\label{Eq_Stern}
(\hat{H}^{(0)}_\mathbf{k}+a\hat{P}_\mathbf{k}-\epsilon_{m\mathbf{k}}^{(0)})
\ket*{u_{m\mathbf{k}}^{\mathcal{E}_\alpha}}=
-\hat{Q}_\mathbf{k}\hat{\mathcal{H}}_\mathbf{k}^{\mathcal{E}_\alpha}
\ket*{u_{m\mathbf{k}}^{(0)}},
\end{equation}
where $\hat{\mathcal{H}}_\mathbf{k}^{\mathcal{E}_\alpha}=i\partial_\alpha\hat{P}_\mathbf{k}+\hat{V}^{\mathcal{E}_\alpha}$, which includes self-consistency via the usual SCF
potential $\hat{V}^{\mathcal{E}_\alpha}$. The tilded first-order wave functions
$\ket*{\tilde{u}_{m\mathbf{k}}^{\mathcal{E}_\alpha}}$ are obtained from 
Eq. (\ref{Eq_Stern}), once the $\hat{V}^{\mathcal{E}_\alpha}$ term is excluded,
\begin{equation}
(\hat{H}_\mathbf{k}^{(0)}+a\hat{P}_\mathbf{k}-\epsilon_{m\mathbf{k}}^{(0)})
\ket*{\tilde{u}_{m\mathbf{k}}^{\mathcal{E}_\alpha}}=
-i\ket*{u_{m\mathbf{k}}^{k_\alpha}}.
\end{equation}
\section{VI. Recovering the formulas of ``Band theory of spatial dispersion in magnetoelectrics" for finite samples}
This section is devoted to establish a formal link between our long-wave DFPT-based expression for the natural optical activity and the formulas obtained in Ref. \cite{PhysRevB.82.245118} for finite systems. First of all, we note that
our \textit{tilded} (see Sec. V) natural optical activity tensor, $\tilde{\eta}_{\alpha\beta\gamma}$, is directly related to the $\mathcal{T}$-even part (where $\mathcal{T}$ represents time-reversal symmetry) of the first $\mathbf{q}$ gradient of the \textit{effective conductivity tensor} computed in Ref. \cite{PhysRevB.82.245118}, $\sigma_{\alpha\beta\gamma}$,
\begin{equation}
\tilde{\eta}_{\alpha\beta\gamma}=\frac{4\pi}{\omega}\sigma^\text{A}_{\alpha\beta\gamma}(\omega)\Big|_{\omega\longrightarrow 0}.
\end{equation}
Our starting point is our Eq. (\ref{Eq_tilde_eta}). In full generality, one can express the first-order wave functions as follows,
\begin{equation}\label{Eq_1_wf}
\ket*{\tilde{u}_{m\mathbf{k}}^{\mathcal{E}_\alpha}}=\sum_{m,l}^{o,e}\ket*{u_{l\mathbf{k}}^{(0)}}
\frac{\bra*{u_{l\mathbf{k}}^{(0)}}\hat{H}_\mathbf{k}^{\mathcal{E}_\alpha}\ket*{u_{m\mathbf{k}}^{(0)}}}{\epsilon_{m\mathbf{k}}^{(0)}-\epsilon_{l\mathbf{k}}^{(0)}},
\end{equation}
where the $m$ and $l$ indices run over the occupied ($o$) and empty ($e$) bands, respectively. For finite samples
\begin{equation}
\hat{H}_\mathbf{k}^{\mathcal{E}_\alpha}\longrightarrow r_\alpha,\quad
\ket*{u_{n\mathbf{k}}^{(0)}}\longrightarrow\ket*{n},\quad
\epsilon_{n\mathbf{k}}^{(0)}\longrightarrow\epsilon_n.
\end{equation}
(Also, and to simplify the notation, we shall use $\hat{H}^{(0)}\rightarrow \hat{H}$.) Along the same lines, it is useful to recall that
\begin{equation}
\partial_\gamma\hat{H}^{(0)}_\mathbf{k}\longrightarrow i[\hat{H},r_\gamma],\quad
\partial_\gamma\hat{P}_\mathbf{k}\longrightarrow i[\hat{P},r_\gamma].
\end{equation}
\subsection{Computing $\tilde{\mathcal{X}}_\mathbf{k}^{\mathcal{E}_\alpha k_\gamma \mathcal{E}_\beta}$}
\begin{equation}
\begin{split}
\tilde{\mathcal{X}}_\mathbf{k}^{\mathcal{E}_\alpha k_\gamma \mathcal{E}_\beta}&=\sum_m^o\bra*{\tilde{u}_{m\mathbf{k}}^{\mathcal{E}_\alpha}}\hat{H}_\mathbf{k}^{k_\gamma}
\ket*{\tilde{u}_{m\mathbf{k}}^{\mathcal{E}_\beta}}\\
&=i\sum_{m,ls}^{o,e}\Big[ 
\frac{\bra*{m}r_\alpha\ket*{l}\bra*{l}[\hat{H},r_\gamma]\ket*{s}\bra*{s}r_\beta\ket*{m}}
{(\epsilon_l-\epsilon_m)(\epsilon_s-\epsilon_m)}
\Big]\\
&=i\sum_{m,ls}^{o,e}\Big[
\frac{\bra*{m}r_\alpha\ket*{l}\bra*{l}\hat{H}r_\gamma\ket*{s}\bra*{s}r_\beta\ket*{m}}
{(\epsilon_l-\epsilon_m)(\epsilon_s-\epsilon_m)}
-\frac{\bra*{m}r_\alpha\ket*{l}\bra*{l}r_\gamma\hat{H}\ket*{s}\bra*{s}r_\beta\ket*{m}}
{(\epsilon_l-\epsilon_m)(\epsilon_s-\epsilon_m)}
\Big]\\
&=i\sum_{m,ls}^{o,e}\Big[
(\epsilon_l-\epsilon_s)\frac{\bra*{m}r_\alpha\ket*{l}\bra*{l}r_\gamma\ket*{s}\bra*{s}r_\beta\ket*{m}}
{(\epsilon_l-\epsilon_m)(\epsilon_s-\epsilon_m)}
\Big]\\
&=i\sum_{m,ls}^{o,e}\Big[
(\epsilon_l-\epsilon_m)\frac{\bra*{m}r_\alpha\ket*{l}\bra*{l}r_\gamma\ket*{s}\bra*{s}r_\beta\ket*{m}}
{(\epsilon_l-\epsilon_m)(\epsilon_s-\epsilon_m)}
+(\epsilon_m-\epsilon_s)\frac{\bra*{m}r_\alpha\ket*{l}\bra*{l}r_\gamma\ket*{s}\bra*{s}r_\beta\ket*{m}}
{(\epsilon_l-\epsilon_m)(\epsilon_s-\epsilon_m)}
\Big]\\
&=i\sum_{m,ls}^{o,e}\Big[
\frac{\bra*{m}r_\alpha\ket*{l}\bra*{l}r_\gamma\ket*{s}\bra*{s}r_\beta\ket*{m}}
{(\epsilon_s-\epsilon_m)}
-\frac{\bra*{m}r_\alpha\ket*{l}\bra*{l}r_\gamma\ket*{s}\bra*{s}r_\beta\ket*{m}}
{(\epsilon_l-\epsilon_m)}
\Big]\\
&=i\sum_{m,l}^{o,e}\Big[
\frac{\bra*{m}r_\alpha\hat{Q}r_\gamma\ket*{l}\bra*{l}r_\beta\ket*{m}}
{(\epsilon_l-\epsilon_m)}
-\frac{\bra*{m}r_\alpha\ket*{l}\bra*{l}r_\gamma\hat{Q}r_\beta\ket*{m}}
{(\epsilon_l-\epsilon_m)}
\Big],
\end{split}
\end{equation}
where $m$ runs over the occupied states $(o)$ only and $s$ and $l$ run over the empty states $(e)$. We have used $\epsilon_l-\epsilon_s=(\epsilon_l-\epsilon_m)+
(\epsilon_m-\epsilon_s)$ in the fifth line and $\hat{Q}=\sum_l\ket*{l}\bra*{l}$. 
\subsection{Computing $\tilde{\mathcal{W}}_\mathbf{k}^{\alpha,\beta\gamma}$}
\begin{equation}\label{Eq_T2}
\begin{split}
\tilde{\mathcal{W}}_\mathbf{k}^{\alpha,\beta\gamma}&=i\sum_m^o
\bra*{\tilde{u}_{m\mathbf{k}}^{\mathcal{E}_\alpha}}
\ket*{u^{A_\beta}_{m\mathbf{k},\gamma}}\\
&=i\sum_{m,l}^{o,e}\frac{\bra*{m}r_\alpha\ket*{l}\bra*{l}\ket*{u_{m\mathbf{k},\gamma}^{A_\beta}}}
{\epsilon_m-\epsilon_l}\\
&=i\sum_{m,l}^{o,e}\frac{\bra*{m}r_\alpha\ket*{l}\bra*{l}\hat{O}^{\beta\gamma}\ket*{m}}
{(\epsilon_m-\epsilon_l)^2},
\end{split}
\end{equation}
where we have used Eq. (\ref{Eq_vetor_pot}) in order to write 
\begin{equation}
\bra*{l}\ket*{u_{m\mathbf{k},\gamma}^{A_\beta}}\longrightarrow
\frac{\bra*{l}\hat{O}^{\beta\gamma}\ket*{m}}{\epsilon_m-\epsilon_l}.
\end{equation}
We shall now focus on computing
\begin{equation}
\bra*{l}\hat{O}^{\beta\gamma}\ket*{m}=
\frac{1}{2}\bra*{l}\big(
\hat{O}^{\text{S},\beta\gamma}+\hat{O}^{\text{A},\beta\gamma}
\big)\ket*{m}.
\end{equation}
\subsubsection{Symmetric part}
\begin{equation}
\begin{split}
\bra*{l}\hat{O}^{\text{S},\beta\gamma}\ket*{m}=&
\underbrace{-\bra*{l}[\hat{H},r_\gamma][\hat{P},r_\beta]\ket*{m}
	-\bra*{l}[\hat{H},r_\beta][\hat{P},r_\gamma]\ket*{m}
	-\bra*{l}[[\hat{H},r_\beta],r_\gamma]\ket*{m}}_{S_1}\\
&\underbrace{+\bra*{l}[\hat{P},r_\beta][\hat{H},r_\gamma]\ket*{m}
	+\bra*{l}[\hat{P},r_\gamma][\hat{H},r_\beta]\ket*{m}}_{S_2}.
\end{split}
\end{equation}
Let us first focus on the $S_1$ term,
\begin{equation}
\begin{split}
S_1=& \underbrace{-\bra*{l}[\hat{H},r_\gamma]r_\beta\ket*{m}
	+\bra*{l}[\hat{H},r_\beta]r_\gamma\ket*{m}
	-\bra*{l}[[\hat{H},r_\beta],r_\gamma]\ket*{m}}_{S_1^{(a)}}
\underbrace{-\bra*{l}[\hat{H},r_\gamma]\hat{P}r_\beta\ket*{m}
	-\bra*{l}[\hat{H},r_\beta]\hat{P}r_\gamma\ket*{m}}_{S_2^{(b)}}.
\end{split}
\end{equation}
The $S_1^{(a)}$ term can be simplified as follows
\begin{equation}
\begin{split}
S_1^{(a)}=&\bra*{l}\hat{H}r_\beta r_\gamma\ket*{m}
-\bra*{l}r_\beta\hat{H}r_\gamma\ket*{m}
+\bra*{l}\hat{H}r_\beta r_\gamma \ket*{m}
-\bra*{l}r_\gamma \hat{H}r_\beta\ket*{m}
-\bra*{l}[\hat{H},r_\beta]r_\gamma\ket*{m}
+\bra*{l}r_\gamma[\hat{H},r_\beta]\ket*{m}\\
=&\bra*{l}\hat{H}r_\gamma r_\beta \ket*{m}
-\bra*{l}r_\gamma r_\beta\ket*{m}\\
=& \bra*{l}[\hat{H},r_\beta r_\gamma]\ket*{m}.
\end{split}
\end{equation}
In the same way, we can simplify the $S_1^{(b)}$ term, where only to terms survive, as we have 
a conduction-band state on the left,
\begin{equation}
\begin{split}
S_1^{(b)}&=\bra*{l}[\hat{P},r_\beta][\hat{H}.r_\gamma]\ket*{m}
+\bra*{l}[\hat{P},r_\gamma][\hat{H},r_\beta]\ket*{m}\\
&=-\bra*{l}r_\beta \hat{P}[\hat{H},r_\gamma]\ket*{m}
-\bra*{l}r_\gamma\hat{P}[\hat{H},r_\beta]\ket*{m}.
\end{split}
\end{equation}
Collecting all terms, $S_1^{(a)} + S_1^{(b)} + S_2$, we obtain
\begin{equation}
\begin{split}
\bra*{l}\hat{O}^{\text{S},\beta\gamma}\ket*{m}=&
\bra*{l}[\hat{H},r_\beta r_\gamma]\ket*{m}
-\bra*{l}[\hat{H},r_\gamma]\hat{P}r_\beta\ket*{m}
-\bra*{l}[\hat{H},r_\beta]\hat{P}r_\gamma\ket*{m}
-\bra*{l}r_\beta \hat{P}[\hat{H},r_\gamma]\ket*{m}
-\bra*{l}r_\gamma\hat{P}[\hat{H},r_\beta]\ket*{m}\\
=&\bra*{l}[\hat{H},r_\beta r_\gamma]\ket*{m}-\bra*{l}\hat{H}r_\gamma\hat{P}r_\beta\ket*{m}
+\Ccancel[red]{\bra*{l}r_\gamma\hat{H}\hat{P}r_\beta\ket*{m}}
-\bra*{l}\hat{H}r_\beta\hat{P}r_\gamma\ket*{m}
+\Ccancel[blue]{\bra*{l}r_\beta\hat{H}\hat{P}r_\gamma\ket*{m} }\\
&-\Ccancel[blue]{ \bra*{l}r_\beta\hat{P}\hat{H}r_\gamma\ket*{m} }
+\bra*{l}r_\beta\hat{P}r_\gamma\hat{H}\ket*{m}
-\Ccancel[red]{r_\gamma \hat{P}\hat{H}r_\beta\ket*{m} }
+\bra*{l}r_\gamma \hat{P}r_\beta\hat{H}\ket*{m}\\
=&\bra*{l}[\hat{H},r_\beta r_\gamma]\ket*{m}-\bra*{l}\hat{H}r_\gamma\hat{P}r_\beta\ket*{m}
-\bra*{l}\hat{H}r_\beta\hat{P}r_\gamma\ket*{m}
+\bra*{l}r_\beta\hat{P}r_\gamma\hat{H}\ket*{m}
+\bra*{l}r_\gamma\hat{P}r_\beta\hat{H}\ket*{m}.
\end{split}
\end{equation}
Thus the symmetric part yields a purely geometric contribution to the wave function response,
\begin{equation}
\bra*{l}\hat{O}^{\text{S},\beta\gamma}\ket*{m}=
(\epsilon_l-\epsilon_m)\bra*{l}\big(
r_\beta r_\gamma -r_\beta\hat{P}r_\gamma -r_\gamma\hat{P}r_\beta
\big)\ket*{m}.
\end{equation}
\subsubsection{Antisymmetric part}
\begin{equation}
\begin{split}
\bra*{l}\hat{O}^{\text{A},\beta\gamma}\ket*{m}&=
\bra*{l}\big(
\partial_\gamma\hat{H}\partial_\beta\hat{P}
-\partial_\beta\hat{H}\partial_\gamma\hat{P}
-\partial_\gamma\hat{P}\partial_\beta\hat{H}
+\partial_\beta\hat{P}\partial_\gamma\hat{H}
\big)\ket*{m}\\
&=\underbrace{-\bra*{l}[\hat{H},r_\gamma][\hat{P},r_\beta]\ket*{m}
	+\bra*{l}[\hat{H},r_\beta][\hat{P},r_\gamma]\ket*{m}}_{A_1}
+\underbrace{\bra*{l}[\hat{P},r_\gamma][\hat{H},r_\beta]\ket*{m}
	-\bra*{l}[\hat{P},r_\beta][\hat{H},r_\gamma]\ket*{m}}_{A_2}.
\end{split}
\end{equation}
Regarding the $A_1$ term, we have
\begin{equation}
\begin{split}
A_1=&-\bra*{l}\big( \hat{H}r_\gamma-r_\gamma\hat{H}\big)
\big(\hat{P}r_\beta-r_\beta\hat{P}\big)\ket*{m}
+\bra*{l}\big(\hat{H}r_\beta-r_\beta\hat{H}\big)
\big(\hat{P}r_\gamma-r_\gamma\hat{P}\big)\ket*{m}\\
=&-\bra*{l}\hat{H}r_\gamma\hat{P}r_\beta\ket*{m}
+\Ccancel[red]{ \bra*{l}\hat{H}r_\gamma r_\beta\ket*{m} }
+\bra*{l}r_\gamma\hat{H}\hat{P}r_\beta\ket*{m}
-\bra*{l}r_\gamma\hat{H}r_\beta\ket*{m}
+\bra*{l}\hat{H}r_\beta\hat{P}r_\gamma\ket*{m}\\
&-\Ccancel[red]{\bra*{l}\hat{H}r_\beta r_\gamma\ket*{m}  }
-\bra*{l}r_\beta\hat{H}\hat{P}r_\gamma\ket*{m}
+\bra*{l}r_\beta \hat{H}r_\gamma\ket*{m}\\
=&-\epsilon_l\bra*{l}r_\gamma\hat{P}r_\beta\ket*{m}
+\bra*{l}r_\gamma\hat{H}\hat{P}r_\beta\ket*{m}
-\bra*{l}r_\gamma\hat{H}r_\beta\ket*{m}
+\epsilon_m\bra*{l}r_\beta\hat{P}r_\gamma\ket*{m}
-\bra*{l}r_\beta\hat{H}\hat{P}r_\gamma\ket*{m}\\
&+\bra*{l}r_\beta\hat{H}r_\gamma\ket*{m}.
\end{split}
\end{equation}
For $A_2$ some terms identically vanish, as we have a conduction-band state on the left,
\begin{equation}
\begin{split}
A_2=&\bra*{l}[\hat{P},r_\gamma][\hat{H},r_\beta]\ket*{m}
-\bra*{l}[\hat{P},r_\beta][\hat{H},r_\gamma]\ket*{m}\\
=&\bra*{l}\big( \hat{P}r_\gamma-r_\gamma\hat{P} \big)
\big(\hat{H}r_\beta-r_\beta\hat{H}\big)\ket*{m}
-\bra*{l}\big( \hat{P}r_\beta-r_\beta\hat{P} \big)
\big( \hat{H}r_\gamma-r_\gamma\hat{H} \big)\ket*{m}\\
=&-\bra*{l}r_\gamma\hat{P}\hat{H}r_\beta\ket*{m}
+\bra*{l}r_\gamma\hat{P}r_\beta\hat{H}\ket*{m}
+\bra*{l}r_\beta\hat{P}\hat{H}r_\gamma\ket*{m}
-\bra*{l}r_\beta\hat{P}r_\gamma \hat{H}\ket*{m}\\
=&-\bra*{l}r_\gamma\hat{P}\hat{H}r_\beta\ket*{m}
+\epsilon_m\bra*{l}r_\gamma \hat{P}r_\beta\ket*{m}
+\bra*{l}r_\beta\hat{P}\hat{H}r_\gamma\ket*{m}
-\epsilon_m \bra*{l}r_\beta \hat{P}r_\gamma\ket*{m}.
\end{split}
\end{equation}
Everything together,
\begin{equation}
\bra*{l}\hat{O}^{\text{A},\beta\gamma}\ket*{m}=\big(\epsilon_l-\epsilon_m\big)
\bra*{l}\big( r_\beta\hat{P}r_\gamma-r_\gamma\hat{P}r_\beta \big)\ket*{m}+
\bra*{l}\big( r_\beta\hat{H}r_\gamma-r_\gamma\hat{H}r_\beta \big)\ket*{m}.
\end{equation}
For convenience, we shall write the second term as
\begin{equation}
\begin{split}
\bra*{l}\big( r_\beta\hat{H}r_\gamma-r_\gamma\hat{H}r_\beta \big)\ket*{m}=&
\bra*{l}\big( r_\beta[\hat{H},r_\gamma]-r_\gamma[\hat{H},r_\beta] \big)\ket*{m}\\
=&-i\bra*{l}\big( r_\beta v_\gamma -r_\gamma v_\beta \big)\ket*{m},
\end{split}
\end{equation}
where we have used $v_\alpha=i[\hat{H},r_\alpha]$. Therefore, the antisymmetric contribution reads
as
\begin{equation}
\bra*{l}\hat{O}^{\text{A},\beta\gamma}\ket*{m}=
-i\bra*{l}\big( r_\beta v_\gamma -r_\gamma v_\beta \big) \ket*{m}
+\big( \epsilon_l-\epsilon_m \big)
\bra*{l}\big( r_\beta\hat{P}r_\gamma-r_\gamma\hat{P}r_\beta \big)\ket*{m}.
\end{equation}
\subsubsection{Symmetric + Antisymmetric contribution}
We can now compute the $\tilde{\mathcal{W}}_\mathbf{k}^{\alpha,\beta\gamma}$ term given by Eq. (\ref{Eq_T2}),
\begin{equation}
\begin{split}
\tilde{\mathcal{W}}_\mathbf{k}^{\alpha,\beta\gamma}=&\frac{i}{2}\sum_{m,l}^{o,e} 
\frac{\bra*{m}r_\alpha\ket*{l}\bra*{l}\big( \hat{O}^{\text{S},\beta\gamma}+
	\hat{O}^{\text{A},\beta\gamma} \big)\ket*{m}}{(\epsilon_m-\epsilon_l)^2}\\
=&\frac{i}{2}\sum_{m,l}^{o,e} \frac{\bra*{m}r_\alpha\ket*{l}}{(\epsilon_m-\epsilon_l)^2}
\Big[
\big(\epsilon_l-\epsilon_m\big)\bra*{l}\big( r_\beta r_\gamma -\Ccancel[red]{r_\beta\hat{P}r_\gamma}
-r_\gamma\hat{P}r_\beta \big)\ket*{m}\\
&+\big(\epsilon_l-\epsilon_m\big)\bra*{l}\big( 
\Ccancel[red]{r_\beta\hat{P}r_\gamma}-r_\gamma\hat{P}r_\beta 
\big)\ket*{m}
-i\bra*{l}\big( r_\beta v_\gamma -r_\gamma v_\beta \big)\ket*{m}
\Big]\\
=&\frac{i}{2}\sum_{m,l}^{o,e} \frac{\bra*{m}r_\alpha\ket*{l}}{(\epsilon_l-\epsilon_m)}
\bra*{l}\big( r_\beta r_\gamma-2r_\gamma\hat{P}r_\beta \big)\ket*{m}
+\frac{1}{2}\sum_{m,l}^{o,e} \frac{\bra*{m}r_\alpha\ket*{l}}{(\epsilon_l-\epsilon_m)^2}
\bra*{l}\big( r_\beta v_\gamma -r_\gamma v_\beta \big)\ket*{m}.
\end{split}
\end{equation}
\subsection{All terms together: computing $\tilde{\eta}_{\alpha\beta\gamma}$}
\begin{equation}
\begin{split}
\tilde{\eta}_{\alpha\beta\gamma}=&-\frac{4\pi}{\Omega}2s\text{ Im}\Big( 
\tilde{\mathcal{X}}_\mathbf{k}^{\mathcal{E}_\alpha k_\gamma\mathcal{E}_\beta}
+\tilde{\mathcal{W}}_\mathbf{k}^{\alpha,\beta\gamma}+
\big[\tilde{\mathcal{W}}_\mathbf{k}^{\beta,\alpha\gamma}\big]^*
\Big)\\
=&-\frac{4\pi}{\Omega}2s\text{ Im}\sum_{m,l}^{o,e}\Bigg\{
i\frac{\bra*{m}r_\alpha\hat{Q}r_\gamma\ket*{l}\bra*{l}r_\beta\ket*{m}}
{(\epsilon_l-\epsilon_m)}
-i\frac{\bra*{m}r_\alpha\ket*{l}\bra*{l}r_\gamma\hat{Q}r_\beta\ket*{m}}
{(\epsilon_l-\epsilon_m)}\\
&\hspace*{2cm}+\frac{i}{2}\frac{\bra*{m}r_\alpha\ket*{l}\bra*{l}\big( r_\beta r_\gamma-2r_\gamma\hat{P}r_\beta \big)\ket*{m}}{(\epsilon_l-\epsilon_m)}
+\frac{1}{2}\frac{\bra*{m}r_\alpha\ket*{l}\bra*{l}\big( r_\beta v_\gamma -r_\gamma v_\beta \big)\ket*{m}}{(\epsilon_l-\epsilon_m)^2}\\
&\hspace*{2cm}-\frac{i}{2}\frac{\bra*{m}\big( r_\alpha r_\gamma-2r_\alpha\hat{P}r_\gamma \big)\ket*{l}\bra*{l}r_\beta\ket*{m}}{(\epsilon_l-\epsilon_m)}
+\frac{1}{2}\frac{\bra*{m}\big( r_\alpha v_\gamma -r_\gamma v_\alpha \big)\ket*{l}
	\bra*{l}r_\beta\ket*{m}}{(\epsilon_l-\epsilon_m)^2}
\Bigg\}.
\end{split}
\end{equation}
We now use $\hat{P}+\hat{Q}=1$ to simplify some terms,
\begin{equation}
\begin{split}
\tilde{\eta}_{\alpha\beta\gamma}=&-\frac{4\pi}{\Omega}2s\text{ Im}\sum_{m,l}^{o,e}
\Bigg\{
\frac{1}{(\epsilon_l-\epsilon_m)^2}\Big[
\bra*{m}r_\alpha\ket*{l}\bra*{l}\big( r_\beta v_\gamma -r_\gamma v_\beta \big)\ket*{m}
+\bra*{m}\big( r_\alpha v_\gamma -r_\gamma v_\alpha \big)\ket*{l}
\bra*{l}r_\beta\ket*{m}
\Big]\\
&\hspace*{2.6cm}+\frac{i}{(\epsilon_l-\epsilon_m)}
\Big[
\bra*{m}r_\alpha r_\gamma \ket*{l}\bra*{l}r_\beta\ket*{m}
-\bra*{m}r_\alpha \ket*{l}\bra*{l}r_\gamma r_\beta\ket*{m}
\Big]
\Bigg\}.
\end{split}
\end{equation}
As a final step, we use the fact that $\text{Im}\big[a-ib\big]=-b$ and 
$\text{Im}\big[i(a-ib)\big]=a$, in order to obtain
the following expression for the natural optical activity tensor for finite samples,
\begin{equation}
\begin{split}
\tilde{\eta}_{\alpha\beta\gamma}&=-\frac{4\pi}{\Omega}s\text{ Im}\sum_{m,l}^{o,e}\Bigg[
\frac{1}{(\epsilon_l-\epsilon_m)^2}\Big(
\bra*{m}r_\alpha\ket*{l}\bra*{l}(r_\beta v_\gamma-r_\gamma v_\beta)\ket*{m}
-\bra*{m}r_\beta\ket*{l}\bra*{l}(r_\alpha v_\gamma-r_\gamma v_\alpha)\ket*{m}
\Big)\\
&\hspace*{2.5cm}+
\frac{i}{(\epsilon_l-\epsilon_m)}\Big(
\bra*{m}r_\beta\ket*{l}\bra*{l}r_\alpha r_\gamma\ket*{m}
-\bra*{m}r_\alpha\ket*{l}\bra*{l}r_\beta r_\gamma\ket*{m}
\Big)
\Bigg],
\end{split}
\end{equation}
which is the same expression that can be obtained from Ref. \cite{PhysRevB.82.245118}.
\clearpage
\section{VII. The two-rank gyration tensor and the optical rotatory parameter for molecules}
The optical rotatory parameter for finite systems like molecules, $\beta$, is usually expressed as the sum of the diagonal components of the two-rank gyration tensor \cite{doi:10.1063/1.2210474,J19710001988}, $g_{\alpha\beta}$, 
\begin{equation}\label{Eq_Rot_power}
\beta=\frac{\Omega}{4\pi}\frac{1}{2}\sum_a\frac{1}{3}g_{aa}.
\end{equation}
In order to go from our three-rank optical activity tensor, $\tilde{\eta}_{\alpha\beta\gamma}$, to
the two-rank gyration tensor, $g_{\alpha\beta}$, all we need to do is to contract one index with
the Levi-Civita symbol (see Eq. (2) of the main text),
\begin{equation}
\begin{split}
g_{\alpha\beta}=&\frac{1}{2}\epsilon_{ab\alpha}\tilde{\eta}_{ab\beta}\\
=&-\frac{1}{2}s\text{ Im}\sum_{m,l}^{o,e}\epsilon_{ab\alpha}
\Bigg[
\frac{1}{(\epsilon_l-\epsilon_m)^2}
\bra*{m}r_a\ket*{l}\bra*{l}\big( r_b v_\beta -r_\beta v_b \big)\ket*{m}
+\frac{i}{(\epsilon_l-\epsilon_m)}
\bra*{m}r_a\ket*{l}\bra*{l}r_b v_\beta\ket*{m}
\Bigg].
\end{split}
\end{equation}
One can then easily obtain the rotatory parameter given by 
Eq. (\ref{Eq_Rot_power}),
\begin{equation}
\begin{split}
\beta=&-\frac{1}{3}s\text{ Im}\sum_{m,l}^{o,e}
\frac{1}{(\epsilon_l-\epsilon_m)^2}\Bigg[
\bra*{m}r_1\ket*{l}\bra*{l}\big( r_2v_3 -r_3v_2 \big)\ket*{m}
+\bra*{m}r_2\ket*{l}\bra*{l}\big( r_3v_1-r_1v_3 \big)\ket*{m}\\
&\hspace*{4.5cm}+\bra*{m}r_3\ket*{l}\bra*{l}
\big( r_1v_2-r_2v_1 \big)\ket*{m}
\Bigg].
\end{split}
\end{equation}
%

%


\end{document}